# Faster and more diverse *de novo* molecular optimization with double-loop reinforcement learning using augmented SMILES

**Esben Jannik Bjerrum\*[1], Christian Margreitter[1], Thomas Blaschke[1], Simona Kolarova[1], Raquel López-Ríos de Castro[1,2]**

1) Odyssey Therapeutics, Cambridge, MA, USA

2) Department of Physics and Department of Chemistry, King's College, London, UK

\*) esben@odysseytx.com

Using generative deep learning models and reinforcement learning together can effectively generate new molecules with desired properties. By employing a multi-objective scoring function, thousands of high-scoring molecules can be generated, making this approach useful for drug discovery and material science. However, the application of these methods can be hindered by computationally expensive or time-consuming scoring procedures, particularly when a large number of function calls are required as feedback in the reinforcement learning optimization. Here, we propose the use of double-loop reinforcement learning with simplified molecular line entry system (SMILES) augmentation to improve the efficiency and speed of the optimization. By adding an inner loop that augments the generated SMILES strings to non-canonical SMILES for use in additional reinforcement learning rounds, we can both reuse the scoring calculations on the molecular level, thereby speeding up the learning process, as well as offer additional protection against mode collapse. We find that employing between 5 and 10 augmentation repetitions is optimal for the scoring functions tested and is further associated with an increased diversity in the generated compounds, improved reproducibility of the sampling runs and the generation of molecules of higher similarity to known ligands.

## Introduction

Deep learning architectures, such as recurrent neural networks (RNNs) trained on simplified molecular line entry system (SMILES) strings were first employed in the generation of molecules that closely resemble the molecular properties of training set compounds more than five years ago.[1]–[3] Since then, SMILES strings have also been used together with generative adversarial networks (GANs)[4] and variational autoencoders[5] in generative applications. Networks that use long short-term memory (LSTM) and are trained to predict the next token or character using teacher's forcing, can efficiently learn the SMILES syntax. This allows them to generate new SMILES strings in a probabilistic manner. These pre-trained networks can then be coupled to optimization algorithms, with early techniques relying on transfer learning and fine-tuning[3], [6], [7], soon followed by methods based on reinforcement learning[2]. Whereas transfer learning needs a dataset of existing ligands to learn the implicit desirable properties, the use of reinforcement learning allows for the flexible definition of molecule scoring functions, which can simultaneously account for a range of explicit and user-defined desirable properties. The original version of REINVENT[2], which coupled pre-trained LSTM networks with reinforcement learning, was developed further with new versions being released[8], along with variants for linker-design[9] and library design[10]. The large array of *de novo* molecular generative algorithms, including SMILES-based methods, has been reviewed previously.[11]–[13]

One of the drawbacks of the use of reinforcement learning in generative algorithms is that it often relies on a large number of scoring function evaluations for the optimization of the agent into a productive state. For instance, in the first REINVENT paper[2] 384,000 scoring function evaluations were used to optimize the agent. This drawback does not present a significant problem if the scoring function has a trivial computational demand, such as that of simple scores like QED, cLogP and fast QSAR models. However, it becomes increasingly problematic as more time- or resource-consuming methods, such as automated molecular docking and scoring, quantum chemical calculations or free energy estimation methods, are used for the scoring function. This issue was recently addressed by *Thomas et al.*, who proposed the use of what they refer to as "augmented hill climb reinforcement learning".[14] It is important to note that the proposed method does not use SMILES augmentation, [15], [16], as the name could imply, and it does not rely on a standard hill-climb algorithm[17], [18], which iteratively fine-tunes the best-scoring compounds sampled. Instead, reinforcement learning is performed on the best-scoring 50% of the compounds in the generated batch of SMILES, combining the augmented loss from the reinforcement learning with the selection of the best-scoring compounds from the hill-climbing. This allows for an increase in the speed of the reinforcement learning; however, the diversity of the generated compounds was not extensively explored in the publication. *Gao et al.* also recently investigated the reinforcement learning efficiency issue by benchmarking multiple generative algorithms.[19] In their publication, REINVENT 2.0 was identified as one of the best-performing algorithms, but only when the sigma parameter, which controls the influence of the scoring function on the loss and the contribution of the loglikelihood of the prior, was increased. Increasing the value of sigma decreases the impact of the prior and may, therefore, increases the risk of over-exploitation, which, in turn, could lead to lower chemical diversity and unrealistic molecules - an issue that the benchmark does not account for and which the authors acknowledge should be improved upon in the future.[20] The issue of optimization of exploration efficiency thus still seems underexplored and unsolved.

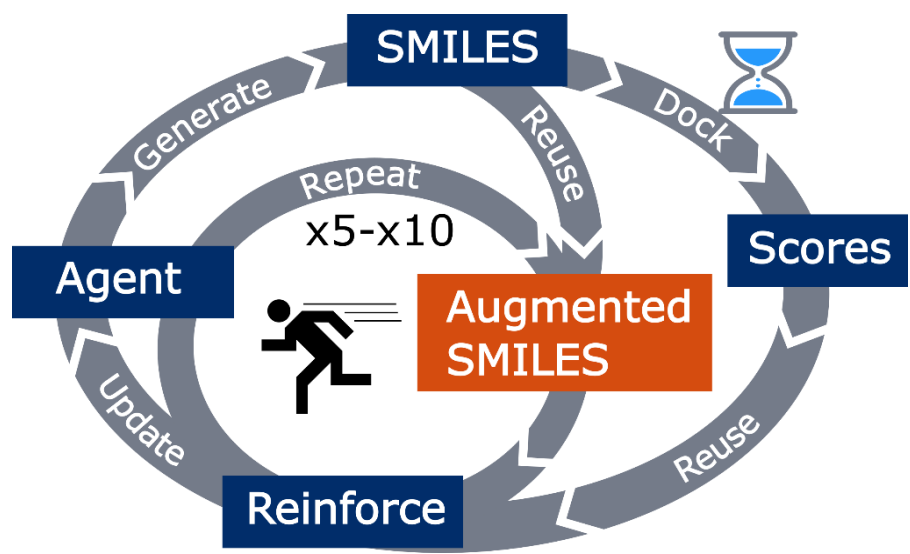

*Figure 1:* Augmented double-loop reinforcement learning. SMILES generated in the outer loop by the agent are augmented and used together with the molecular scores produced in the outer loop to further reinforce the agent in the inner loop. This allows for the agent to be updated several using molecules scored only once.

Here, we present a method for increasing the speed of the reinforcement learning optimization, which utilizes SMILES augmentation to provide more efficient feedback to the agent and speed up learning and exploration. SMILES augmentation, also called SMILES enumeration[15] or SMILES randomization[16], is the process of using several different non-canonical SMILES representations of the same molecule with the purpose of improving deep learning. The use of SMILES augmentation has been shown to improve task performance and accuracy for a range of deep learning algorithms in the areas of deep learning QSAR[15], molecular generation[16], [21], latent space optimization[22], unsupervised pretraining[23] and retrosynthesis prediction[24].

First, we explore how the number of augmentations affects reinforcement learning optimization in an exploitative setting with a similarity-based scoring function targeting a single reference molecule. We compare it to similar increases in the learning rate to rule out the possibility that the improved performance is only due to the consecutive smaller inner loop updates resulting in larger weight updates. Then, we continue by examining the optimization in an exploratory setting with a docking target. We examine not only the speed of obtaining larger docking scores for generated molecules, but also characterize the diversity of the generated molecules using internal diversity[25], yield curves, and ChemCharts[26] visualizations. We further compare the algorithm's efficiency in re-finding compounds similar to known ligands using solely docking as the scoring method. Lastly, we compare how augmentations enhance two different variants of reinforcement algorithms, REINVENT 2.0 and augmented hill-climb[14], using a QSAR model as the scoring task.

Overall, we show that by combining SMILES augmentation with reinforcement learning, we increase both the efficiency and the molecular diversity and reproducibility of the molecular optimization process.

## Methods

REINVENT 2.0[8] was forked and its reinforcement learning loop was altered by the inclusion of an inner loop (Figure 1). In this inner loop, alternative SMILES sequences are obtained for the SMILES generated by the outer loop's agent via SMILES augmentation[15] and are then used together with the molecular scores produced in the outer loop to further reinforce the agent. The number of augmentations in the inner loop is a configurable hyperparameter, which decouples the agent's updates via reinforcement learning from slow-scoring components, such as docking Since the generation of augmented SMILES and the backpropagation in the reinforcement learning loop are both considerably faster processes than the scoring of molecules, the overall procedure can be carried out faster and requires fewer scoring function evaluations. We named the updated REINVENT version *GenAI*.

### Similarity as exploitation task

To assess the performance of *GenAI*, an entirely exploitative task was carried out first. The aim of the task was to optimize the fingerprint-based Tanimoto similarity of generated compounds to a target compound, namely Aripiprazole selected from the GuacaMol benchmark[18] (SMILES: "Clc4cccc(N3CCN(CCCCOc2ccc1c(NC(=O)CC1)c2)CC3)c4Cl"). For this purpose, the similarity scoring component of *GenAI* was configured with the target SMILES and given a weight of one; the default fingerprint was utilized (unhashed counted Feature Morgan fingerprint with a radius of 3); experience replay with a memory size of 100 and a sample size of 10 was used; and no diversity filter was employed[27]. The batch size for generation was set to 64. Sigma was kept as the default value of 128 for all runs. The prior used, *i.e.*, "random.prior.new", was obtained from the REINVENT Community repository.[28]

Two sets of experiments were initially carried out to separately assess the effects of the augmentation and the learning rate hyperparameters. In the former, the number of augmentations was varied in the range of 1, 2, 5, 10, 20, 50, 100 and 200, while the learning rate was fixed at the default value of 0.0001; in the latter, the number of augmentations was fixed to 1, while the learning rate was varied by multiplying the default value of 0.0001 with a value in the range of 1, 2, 5, 10, 20, 40, 60, 80, and 100, resulting in learning rates between 0.0001 and 0.01. A total of 1000 steps were used for each run, corresponding to roughly 64,000 scoring function evaluations. Subsequently, a grid search was used to simultaneously vary the number of augmentations (1, 2, 3, 5, 10, 20) and the learning rate multiplication factor (1, 2, 3, 5, 7, 15) using 500 steps. All experiments were conducted in triplicates.

### Docking as exploration task

The second task was designed to assess the diversity of the molecules produced with *GenAI* by generating inhibitors for a dopamine type 2 receptor (D2R) docking target (gene symbol DRD2). For this purpose, the protein structural file (PDB ID: 6CM4[29]) was obtained from the Protein Data Bank (PDB)[30] and prepared for docking using Schrodinger's Glide software[31]. The docking scoring component was set up using DockStream[32] with a Glide backend[31]. Specifically, a reverse sigmoid transformation between -1 and -11 and a *k* parameter of 0.25 were used for the DockStream constituent. The molecular weight of the generated compounds was restricted via a second scoring component with a double sigmoid transformation between 0 and 500 Da. The two scoring components were each assigned a weight of one and their product was used as the final score.

As in the similarity task, the batch size was 64, sigma was 128, and experience replay with a memory size of 100, and a sample size of 10 was used. Notably, a diversity filter[27] based on the Bemis-Murcko scaffold similarity with a bin size of 35, a minimum similarity of 0.4 and a minimum score of 0.5 was also employed. This diversity filter penalized the agent by setting the score for a generated compound to zero if either the molecule has been previously sampled or if compounds with the same scaffold have already been generated more than a set number of times, *i.e.*, 35. As above, two sets of experiments were carried out to assess the effect of the number of augmentation and the learning rate used: in the former the augmentations hyperparameter was varied in the range of 1, 2, 5, 10, 25, 50 and 100, while in the latter the multiplication factor for the default learning rate (0.0001) was varied in the range of 1, 2, 5, 10, 25, 50 and 100. A total of 500 steps were used for each run (roughly 32,000 scoring function evaluations) and all experiments were performed in triplicates.

### Algorithm comparison with QSAR task

Finally, the ability of SMILES augmentation to enhance the performance of alternative generative algorithms was assessed. For this purpose, four algorithms were compared: the REINVENT 2.0 algorithm; the REINVENT 2.0 algorithm with an inner SMILES augmentation loop of five repeats as implemented in *GenAI* (referred to as *GenAI 5xA* below); an augmented hill-climb (*AHC*) algorithm implemented in *GenAI* following the methodology described by *Thomas et al.*[14]; and the same AHC algorithm with the additional use of an inner loop of 5 repeats, in which the SMILES of the top-k scoring compounds obtained *via* generation were augmented and used for reinforcement learning (referred to as *AHC 5xA* from now on). The top-k parameter describing the fraction of generated molecules whose score is included in the loss and backpropagated to the agent was set to 0.5 in both AHC algorithms.

All four algorithms were set up for the same goal – to generate D2R-binding molecules by optimizing a D2R QSAR model's predicted affinity score. The QSAR model was trained on molecules with a reported D2R binding affinity in the ExCAPE-DB[33] database (DRD2 gene symbol). All compounds with an "A" flag and pXC50 values larger than 5 were considered to be binders and used in model training, while the remaining compounds with a "N" flag were considered to be non-binders and a random selection of 100,000 of them was used in model training. Stereo information was removed from the molecules and duplicates removed from the dataset. This dataset was randomly split into a training and test set with a 3:1 ratio resulting in 81,188 molecules in the final training set and 27,063 molecules in the test set. The machine learning algorithm used was a C-support vector classification from scikit-learn (version 1.0.2)[34] trained on count-based extended connectivity fingerprints (ECFPs) with 2048 bits and a radius of 3 atoms calculated using the Morgan fingerprint algorithm in RDKit.[35] Model hyperparameters were tuned on the training data subset using scikit-learn's randomized search (RandomizedSearchCV) and a three-fold cross-validation over 50 iterations. The *gamma* and *C* parameters were varied between $10^{-5}$ and $10^{0}$ and between $10^{-1}$ and $10^{3}$, respectively. The optimum hyperparameter values obtained from the tuning were 29.310978 and 0.00652 for *C* and *gamma*, respectively. These hyperparameters were then used to retrain the model on the entire training set and the model's predictive performance was assessed on the external test set which gave an F1 score of 0.99, a Matthew's correlation coefficient of 0.96 and a balanced accuracy of 0.98.

The generative runs for all four generative algorithm variants were performed using identical settings. Specifically, no inception or experience replay was used, and a Bemis-Murcko scaffold diversity filter with a bin size of 25, a minimum similarity of 0.4 and a minimum score of 0.4 was employed. The scoring function was composed of two elements: the QSAR model described above with a weight of 6 and a molecular weight restraint, which was modelled via a double sigmoid transformation with an upper limit of 620 Da and a lower limit of 180 Da, with a weight of 4. The batch size was set to 128 SMILES, the learning rate - to the default value of 0.0001 and sigma - to the default value of 128. The margin threshold was fixed to 50 and, as above, the prior "random.prior.new" obtained from the REINVENT Community repository[28] was used. All experiments were run for 500 steps in triplicate.

### Analysis of generative runs

To assess the performance of the generative runs, the average score, the validity of the generated SMILES, as well as the number of molecules with a score higher than the minimum score (0.5 or 0.4) was logged and plotted. For the similarity tasks, the logs were further analyzed to determine the first step at which the similarity surpassed thresholds of 0.5 or 0.8 or the target compound was generated. Additionally, the yield of the generative runs, *i.e.*, the number of unique and valid well-scoring compounds generated expressed as a fraction of the total number of generated SMILES strings, was determined at different scores and plotted. The internal diversity of the generated unique molecules with a score higher than 0.5 was calculated using the IntDiv1 metric from the Moses benchmark.[25] The internal diversity is calculated based on the average Tanimoto fingerprint similarity between the compounds in a dataset and is a quantity that characterizes a single dataset rather than the similarity between two datasets, *e.g.*, the training dataset and generated compounds, as is often the case in other generative metrics.[18], [25]. For comparison purposes, the internal diversity of the D2R binders from ExCAPE-DB[33] was also calculated after removing stereochemistry and filtering duplicate compounds.

The chemical space coverage was visually analyzed using Chemcharts[26] with standard settings. The triplicate runs for each generative algorithm were pooled before dimensionality reduction and plotting. ChemCharts creates a joint UMAP[36] embedding of fingerprints (RDKfingerprint) and the individual datasets were plotted in hex-binned charts. The embeddings of the samples were further used to calculate the cosine similarity between the datasets.

# Results

## Assessment using similarity task

With the aim of assessing the generative performance of *GenAI* and the effects of different hyperparameters on it, the results of the similarity task – a purely exploitative task designed to optimize the fingerprint-based Tanimoto similarity of generated compounds to a target compound, *i.e.*, Aripiprazole, are presented here. Specifically, the influence of variations in the number of inner-loop SMILES augmentation and learning rate were first analyzed independently and subsequently together below.

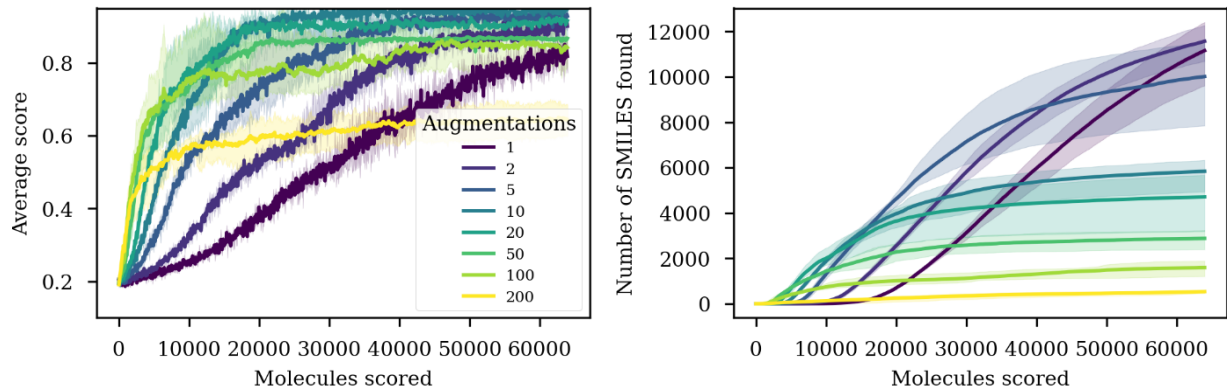

*Figure 2: Effect of the number of SMILES augmentations on the generative efficiency of GenAI. LEFT: Relationship between the number of inner-loop SMILES augmentations and the average similarity score achieved at different points in the generative run. Increasing the number of inner-loop SMILES augmentations moderately is correlated with an increase in the efficiency of the search. RIGHT: Relationship between the number of inner-loop SMILES augmentations and the number of unique molecules sampled with a score over 0.5. A higher number of augmentations is correlated with an earlier increase in the number of unique SMILES generated. Curves are the mean of triplicate experimental runs with the shaded area encompassing all obtained results.*

First, the effect of the number of SMILES augmentations in the inner loop of *GenAI* on the generative performance for the similarity task was assessed. The results shown in Figure 2 (left panel) suggest that a moderate increase in the number of augmentations is related to a marked increase in the efficiency of the generation, as evidenced by the reduced number of molecule evaluations required to obtain highly scoring, *i.e.*, highly similar to Aripiprazole, molecules. Nevertheless, using the highest numbers of augmentations studied here, 100 or 200, leads to a lower maximum score obtained at the plateau. These results suggest that the optimum number of inner-loop augmentations is 10, resulting in both a high generative efficiency, as well as a high maximum similarity score at the plateau. A similar trend is also reflected in Figure 2 (right panel), which shows the relationship between the number of SMILES augmentations and the number of SMILES surpassing a similarity score threshold of 0.5. Specifically, increasing the number of augmentations leads to a faster rise in the yield of SMILES surpassing the scoring threshold; however, it is also correlated with the faster reaching of a plateau, which is indicative of mode collapse, *i.e.*, the same highly scoring SMILES are repeatedly generated. From this perspective, the optimum number of inner-loop augmentations depends on the intended number of scored molecules, *i.e.*, the available computational time and resources of the generative run. For instance, if the early return of some high-scoring SMILES is required (*e.g.*, 10,000 molecules are to be scored as in the proposed efficiency benchmark[19]), then the optimum number of SMILES augmentations would be around 10; whereas if more computational time and resources can be used, then a lower number of augmentations would be preferential. It is important to note that determining the optimum number of inner-loop SMILES augmentations using the results of a diversity-oriented metric, such as the yield of SMILES, on experimental runs configured for exploitation, might not be optimal, as the metric is more suited to address the performance of explorative tasks. Additional information about the validity of the generated SMILES is provided in the Supplementary Material (Supplementary Figure 1) with the results suggesting that the percentage of valid SMILES generated is in the high nineties and appears to increase with the number of augmentations.

To further gauge the effect of the inner-loop SMILES augmentations on generative performance, the relationship between the number of augmentations and the number of steps required to solve the similarity task was investigated and reported in Table 1 (upper half). In particular, three different definitions of a solved task were applied: the target compound is generated; the generated compounds exceed an average score of 0.8; and the generated compounds exceed an average score of 0.5. In general, increasing the number of augmentations leads to less steps needed before passing the thresholds or sampling the target compound. In fact, the fastest generative run to re-find the target compound in only 73 steps employed 100 augmentations. However, as is evident from the table, the highest number of augmentations lead to instabilities in the runs, as more runs are not passing the threshold or finding the target compound within 1000 steps (unsolved or star-marked). As above, the optimum number of inner-loop SMILES augmentations depends on the generative run scenario and the available computational time and resources (marked with bold for each column). Specifically, if medium scoring compounds are required (meeting the lowest solution criterium of an average score higher than 0.5) and the computational resources are limited, then the optimum strategy would be to employ a relatively large number of augmentations; whereas if higher scoring compounds are wanted (with an average score higher than 0.8 or comprising the target compound), employing a more moderate number of augmentations (around 10) is optimal.

*Table 1: Number of steps required to solve the similarity task by meeting one of three criteria: the generated compounds exceed an average similarity score of 0.5, the generated compounds exceed an average similarity score of 0.8 and the target compound is generated. The mean number of steps required +/- the standard deviation of three runs is reported. If a solution criterium was not met by the end of an experiment for one, two or all three of the total of three conducted experimental runs, this is denoted with a \*, a \*\* or "Unsolved", respectively. Run numbers in bold show the lowest mean steps required to meet each criterion for the augmentation and the learning rate searches separately, disregarding runs that were not successful in all three runs. Experiments were performed over 1000 steps.*

| Learning rate multiplication | Number of Augmentations | Similarity >0.5 | Similarity >0.8 | Target Sampled |
|---|---|---|---|---|
| 1 | 1 | 414 ± 27 | 781 ± 19 | 473 ± 178 |
| 1 | 2 | 260 ± 7 | 550 ± 16 | 320 ± 62 |
| 1 | 5 | 133 ± 4 | 375 ± 75 | 242 ± 73 |
| 1 | 10 | 93 ± 9 | 194 ± 31 | **131 ± 38** |

| 1  | 20  | 64 ± 2   | **173 ± 57** | 92*      |
| 1  | 50  | 40 ± 4   | 222 ± 43     | Unsolved |
| 1  | 100 | **39 ± 11** | 427 ± 239 | 73**     |
| 1  | 200 | 63 ± 35  | Unsolved     | Unsolved |
| 2  | 1   | 251 ± 12 | 535 ± 137    | **214 ± 72** |
| 5  | 1   | 140 ± 6  | 444 ± 33     | 593*     |
| 10 | 1   | 103 ± 9  | **276 ± 69** | 365 ± 338 |
| 20 | 1   | 90 ± 2   | 348 ± 34     | 275**    |
| 40 | 1   | **67 ± 10** | 237**     | 84**     |
| 60 | 1   | 95 ± 9   | Unsolved     | Unsolved |
| 80 | 1   | 113*     | Unsolved     | Unsolved |

Next, the effect of the learning rate on the generative performance of *GenAI* for the similarity task was assessed by conducting similarity search runs with different learning rate multiplication factors. The results in Figure 3 (left panel) suggest that increasing the learning rate multiplication factor from 1 to 10 enhances the ability of the generative algorithm to obtain highly scoring compounds; however, similarly to the effect of the number of inner-loop augmentations on generation discussed above, there appears to be an optimum learning rate multiplication factor around 10 above which the average score plateaus at relatively lower average score values (i.e., a plateau at ~0.65 for x100 compared to ~0.9 for x10). Notably, the variations in the average scores of the runs are larger for higher learning rates. Higher learning rates also give less reproducible runs as can be seen in the large number of runs that fail to find a solution to the similarity task, as evidenced in Table 1 (lower half). This instability can also be observed in the results in Figure 4, which show that the SMILES validity at the beginning of the generative runs is markedly low when the highest learning rates are employed but surprisingly recovers as the agent learns to generate valid SMILES even under the extreme neural network weight update conditions. Interestingly, the results in Figure 3 (right panel) suggest that the number of unique SMILES generated is not as strongly affected by changes in the learning rate as it is by changes in the number of inner-loop augmentation (see above); nevertheless, the variability in yield is relatively high, especially at high learning rates.

To further understand whether the improvements observed with the addition of inner-loop SMILES augmentations are simply a result of an increase in the updating rate of the agent's weights, *i.e.*, an effective increase in the learning rate obtained with multiple smaller updates, a number of generative runs with the default learning rate and several inner-loop repeats reusing the sampled SMILES sequences without augmenting them were conducted. The results of these runs are plotted in Supplementary Figure 4 and suggest that the presence of an inner loop of several repeats itself does influence the generative efficiency; however, these runs had a considerably higher propensity for reaching a plateau of suboptimal average score than the runs in which SMILES augmentation was employed. Additionally, even with just a single non-augmenting step in the inner loop, the yield of SMILES is lower than that for the runs utilizing SMILES augmentation (shown in Figure 2). Moreover, generative runs employing as little as three augmentations appear to be more efficient when SMILES augmentation is used (Results not shown for three augmentations). Notably, in some cases the efficiency of the non-augmented inner-loop runs can be higher than that of the augmented inner-loop generative runs as is the case when a single inner loop is employed (Supplementary Figure 4 and Figure 2). Thus, it can be ruled out that the increase in efficiency using double-loop augmented reinforcement learning is solely caused by too many small weight updates.

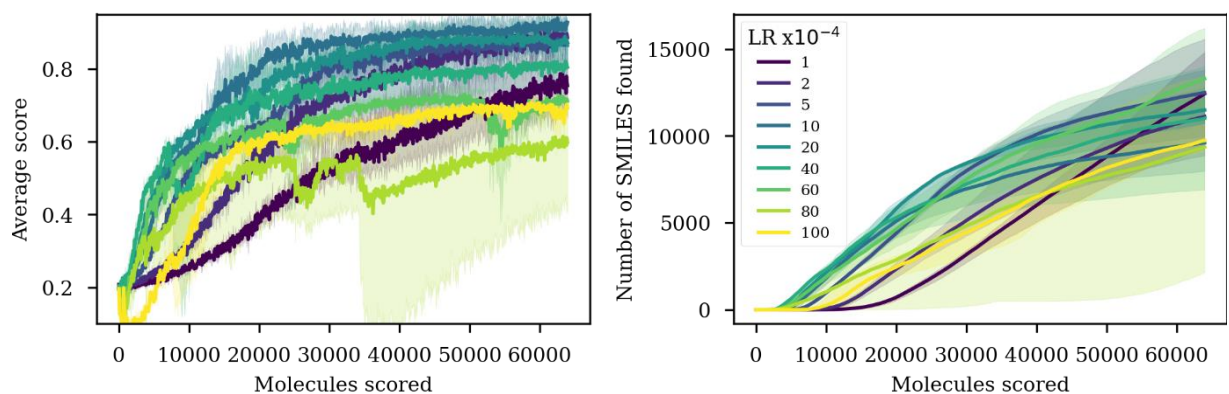

Figure 3: Effect of the learning rate multiplication factor on the generative performance of GenAI for the similarity task. LEFT: Effect on the efficiency of score optimization. A higher learning rate leads to obtaining good scores faster, but a too high learning rate leads to instability. RIGHT: Effect on the number of generated SMILES with a score higher than a threshold of 0.5.

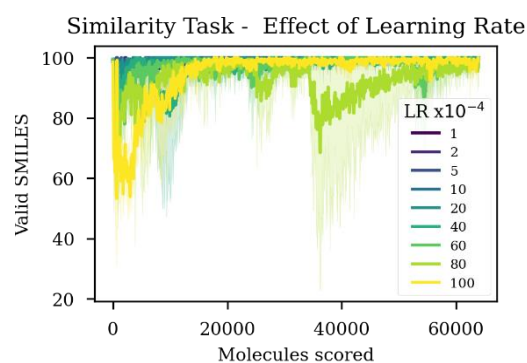

Figure 4: Effect of learning rate multiplication factor on SMILES validity. The highest settings lead to instability and a lower fraction of valid SMILES. Surprisingly, the agent learns to recover from this.

Finally, a grid search experiment was conducted in which both the number of inner-loop SMILES augmentations and the learning rate multiplication factor were varied. The heatmap in Figure 5 (left panel) depicts the effect of these variations on the number of steps required for the average similarity score to exceed 0.5 and highlights the synergy between increasing the learning rate and increasing the number of augmentations for improving the generative efficiency. Nevertheless, as shown in Figure 5 (right panel), the risk of unsuccessful runs quickly increases when combining a higher learning rate with a higher number of augmentations. The numerical results of the grid search are shown fully in Supplementary Table 1. The generative run setup for which a solution was obtained, *i.e.*, the Aripiprazole molecule was generated, for all three experimental runs in the least average number of steps, *i.e.*, 122, utilized two inner-loop augmentations and a learning rate multiplication factor of three. The fastest single generative run that generated the target compound in only 44 steps had an augmentation number of five in combination with a learning rate multiplication factor of seven; however, only one out of three runs with these hyperparameters found the solution in 500 steps.

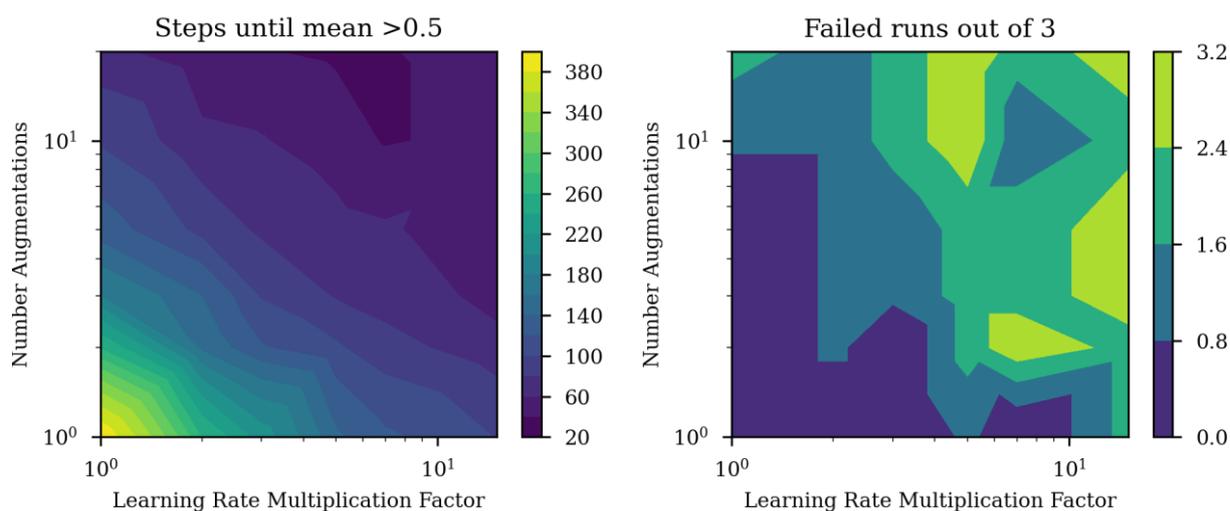

Figure 5: *Effect of simultaneously varying the learning rate and the number of augmentations. LEFT: Effect on the number of steps required to obtain an average score higher than 0.5. RIGHT: Effect on the number of runs that fail to find the target compound across three generative runs.*

Overall, both moderately increasing the learning rate and performing multiple rounds of augmentation improve the optimization efficiency, reducing the number of molecular scorings necessary to reach good scores. However, using large values for these hyperparameters can lead to instability and failure to reach the target score or compound. Increasing the learning rate is more sensitive to this issue compared to increasing the number of augmentations.

### Assessment using docking task

The second task was an explorative task utilizing automated docking and scoring for a dopamine type 2 receptor docking target (D2R) receptor with the aim of generating a large number of diverse molecules with good docking scores. Notably, a molecular weight penalty was added to the compound scoring function and a diversity filter was utilized during the generative runs to restrain the size of the molecules to a reasonable range and encourage exploration over exploitation, respectively. Likewise to the similarity task, the influence of altering the number of inner-loop augmentations and the learning rate on the generative efficiency and the SMILES yield is assessed below and, additionality, the diversity of the generated compounds is also analyzed.

First, the outcome of the docking task for different numbers of augmentations is shown in Figure 6. Similarly to the previous task, the number of augmentations is also observed to influence the effectiveness with which the average docking score increases as molecules are scored; however, unlike for the similarity task, the efficiency, as well as the average score at the plateau, appear to be optimum for the largest number of augmentations explored, *i.e.*, 100 (Figure 6 (left panel)). Interestingly, some of the curves in Figure 6 (left panel) exhibit a spike in the average score at the beginning of the generation with the score rapidly increasing and then slightly dropping but recovering in the long run. We attribute this to the influence of the diversity filter that controls overexploitation. Even though the average score appears to be highest when 100 augmentations are employed, the plot in Figure 6 (right panel) suggests that the yield of SMILES for this number of augmentations is suboptimal. Instead, the yield of unique SMILES surpassing the scoring threshold of 0.5 is highest when 5 or 10 augmentations are used with the latter augmentations number resulting in higher results variability. The SMILES validity percentage was recorded to be in the high nineties for all runs (Supplementary Figure 2).

Variations in the learning rate also influence the efficiency of the reinforcement learning for the docking task. Specifically, higher learning rates allow the reinforcement learning to find better scoring molecules faster for all learning rate increases except for the highest one (Figure 7 (left panel)). Interestingly, the generative run with the highest learning rate multiplication factor, *i.e.*, 100, also exhibits a large variation in the average score recorded for the three experiments, which may in part be correlated with the instability of the runs and the marked decrease in SMILES validity (Supplementary Figure 3) that was also previously observed for similarity task runs of the same learning rates (see above). In contrast to the findings for the learning rate variations with no augmentations (*cf.* Figure 6 (right panel)), increasing the learning rate affected the final yield of SMILES negatively across all learning rate increases (Figure 7 (right panel)) regardless of the improvements observed at intermediate steps of the runs.

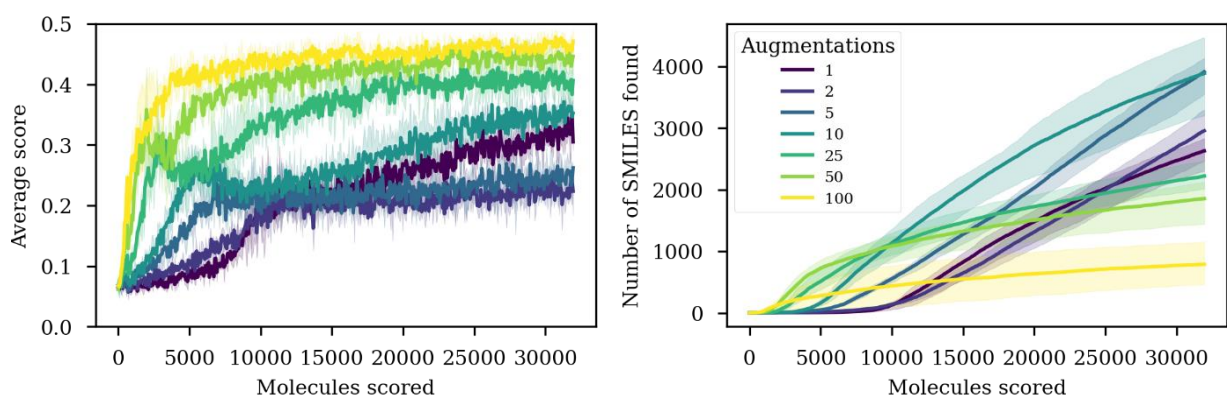

*Figure 6: Effect of the number of SMILES augmentations in the Docking task. LEFT: Effect on optimization efficiency. More augmentations lead to higher optimization efficiency. RIGHT: Effect on the yield of unique SMILES with a score above a threshold of 0.5*

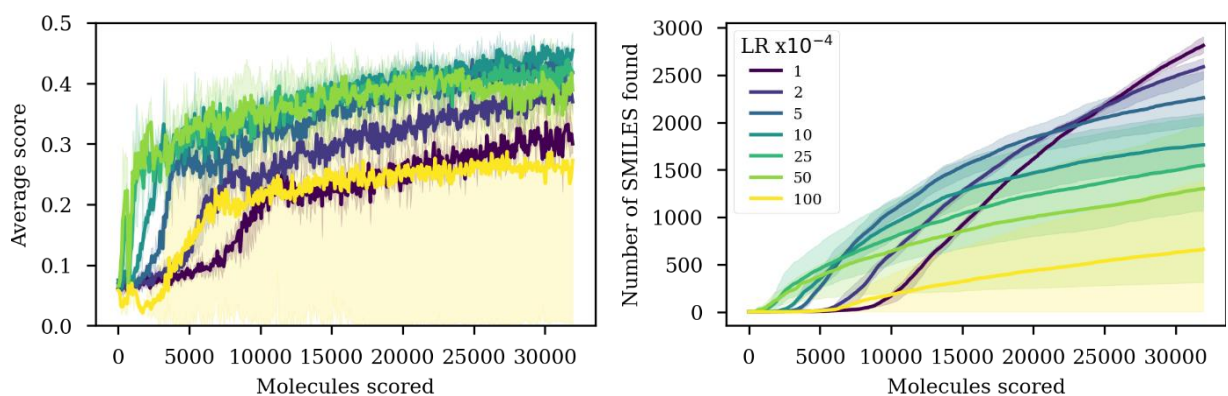

*Figure 7: Effect of the learning rate multiplication factor on the Docking task. LEFT: Medium learning rate increases leads to higher optimization efficiency. RIGHT: The number of SMILES found initially increases with higher learning rate, but all increases lead to a lower number of obtained unique SMILES at the end.*

### Yield and diversity of the generated compounds

Since researchers aim to generate both highly scoring compounds and a diverse selection of compounds for post-processing, this section will discuss the diversity and yield of the generative process, as well as the chemical space coverage of the generated compounds. The yield curves in Figure 8 and Figure 9 show the changes in yield fraction with different score thresholds for the series of experiments in which the number of augmentations and the learning rate were varied, respectively. The plots suggest that as the learning rate increases the yield fraction decreases over the entire range of score thresholds, whereas as the augmentations number increases the yield also initially increases (for runs with 1 to 10 augmentation loops) before decreasing (for runs with 25 to 100 augmentation loops).

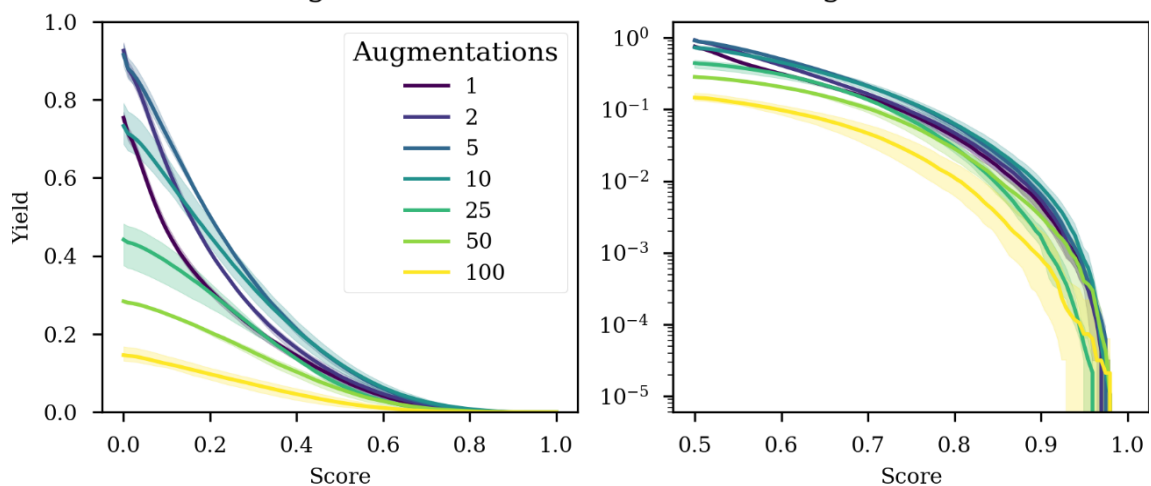

Figure 8 Yield fraction curves for the augmented runs of the docking task. Left shows the total yield curves, whereas the right shows the compounds scoring over 0.5 on a semi-logarithmic plot. For 2-5 augmentations the yields are highest for most score thresholds, but the yield drops with a higher number of augmentations.

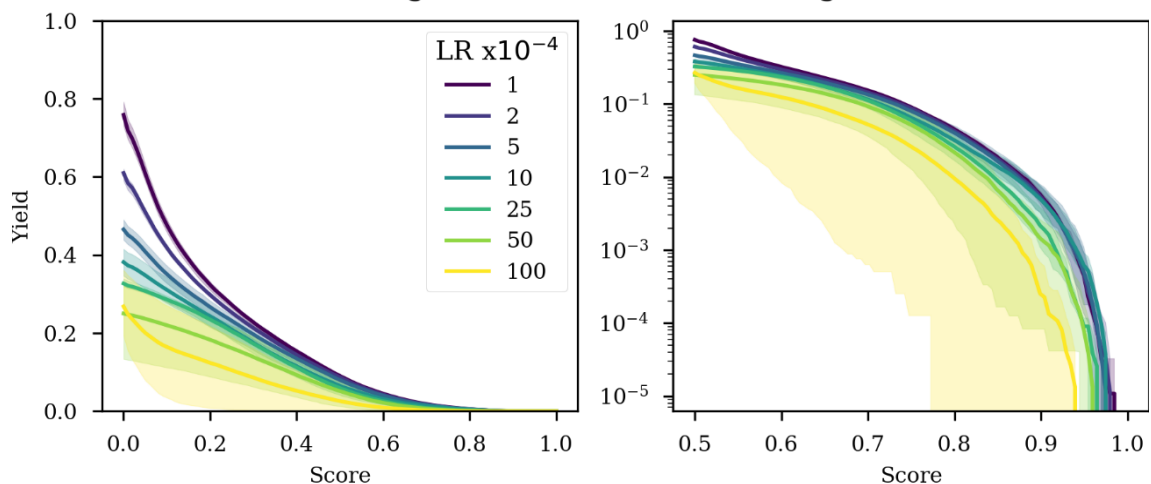

Figure 9 Yield fraction curves for the runs with varying learning rates. Left shows yield for all scores and right shows the number of compounds scoring over 0.5 in a semi-logarithmic plot. Curves are the means of triplicate runs with min and max as the shaded area. Yield drops with increasing learning rate across the entire range of scores and leads to higher instability for largest increases.

Next, the diversity of the generated compounds was quantified using the internal diversity measure from the Moses benchmark[25] (IntDiv1) and compared to the internal diversity of the D2R binders. Figure 10 shows the variations in the internal diversity with different configurations of hyperparameters. The plot suggests that higher augmentation numbers generally improve the internal diversity of the generated compounds with a maximum achieved at five augmentations (with the exception of runs performed with 100 augmentations, for which the greatest results variability is also observed). In contrast, an increase in the learning rate leads to a less notable increase in internal diversity but a higher variation between triplicate runs, especially for the higher learning rate explored. The IntDiv2 metric showed a qualitatively comparable profile to the internal diversity albeit with lower absolute diversity values (results not shown).

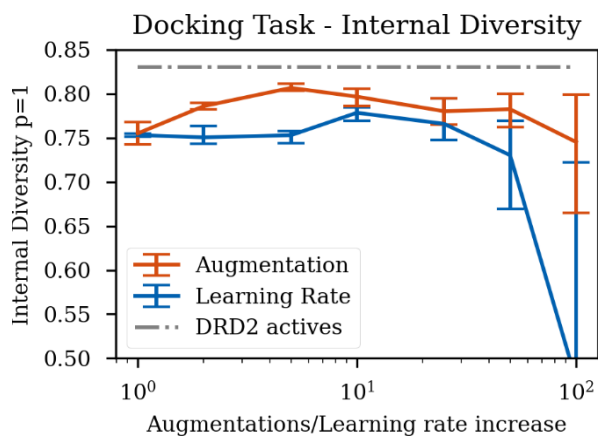

*Figure 10: Mean internal diversity of the docking task generative runs. Error bars are min and max of the triplicates.*

The coverage of chemical space by the generated compounds was also examined visually using the ChemCharts[26] package. By employing UMAP[36] to define a common 2D space for all experimental runs, the distributions of the generated compounds were visualized in the 2D hexbin plots shown in Figure 12 and Figure 13 for the experiments in which the number of inner-loop augmentations and the learning rate were altered, respectively. Interestingly, generative runs utilizing a moderate number of augmentations (between 2 and 10) appear to sample a similar chemical space area. Moreover, these runs also seem to produce a relatively spread-out molecular sampling, which agrees well with the observation above regarding the internal diversity of the generated compounds (Figure 10). In contrast, raising the learning rate appears to gradually lead to partial mode collapse and the narrower focusing of the generated compounds in one or a few areas of the chemical space (Figure 13).

Finally, the similarity between all datasets of generated compounds was compared using the cosine similarity calculations provided in the ChemCharts package and is shown in the heatmap in Figure 14. As a baseline, the similarity between the two pooled generative run triplicates denoted *Aug1* and *Lr1* in Figure 14 was first calculated. These runs were conducted using identical default settings, *i.e.,* a single inner-loop augmentation and a learning rate of 0.001, and were determined to have a cosine similarity of 0.89 that suggests a largely comparable chemical space was sampled. Nevertheless, the similarity between the generative runs employing 2, 5 and 10 SMILES augmentations was recorded to be higher (0.90 – 0.97), which indicates that the use of inner-loop augmentations increases the reproducibility of the generated chemical space - a surprising observation in view of the reported above highest internal diversity for these generated sets (*cf.* Figure 10 and Figure 12). Notably, the similarity of the runs using between 2 and 10 augmentation loops to the runs using a single augmentation is lower (0.64 – 0.68), which indicates that these runs are sampling a slightly different chemical space or sampling some areas to a different degree. The previously discussed results in Figure 12 (top row) seem to support the latter suggestion as the overall spread in chemical space is similar but intensities appear to vary. As for the influence of the learning rate on these results, we generally observe a gradual diminishing in similarity between molecules generated using runs of lower *vs.* higher learning rates as the learning rate of either run is increased. In other words, experimental runs of relatively smaller learning rates seem to be most like each other, while these of relatively higher learning rate appear to be much less reproducible. Notably, increasing either the learning rate multiplication factor or the number of augmentations to the highest studied values, *i.e.*, 100, produced results that can be considered outliers with respect to the cosine distances to the rest of the datasets, which can likely be attributed to mode collapse – a suggestion that is also supported by the plots in Figure 12 and Figure 13 (bottom rows).

Overall, the assessment of the effects of learning rate and number of augmentations on the docking tasks follow similar lines as the effects observed for the exploitive similarity task. Moderate increases in both are seemingly beneficial for the optimization efficiency, while high increases lead to diminished yields as well as a lower final average scores. Increases in learning rates seem more prone to bringing about these effects and, for example, the yield immediately starts to drop upon increasing the learning rate, whereas it increases with the number of augmentations up to 5-10 augmentations. The effects on internal diversity also suggest that a moderate number of augmentations are beneficial, whereas increasing the learning rates does not result in as pronounced or stable internal diversity improvements. Moreover, by comparing the cosine similarities of the sampled datasets it can be deduced that moderate increases in augmentations (2 to 10) lead to a higher similarity between generative runs, indicative of more reproducible runs. We conclude that a number of augmentations around five seems most beneficial for this task.

### Similarity to known D2R ligands

Since the docking runs generated compounds using a scoring function that did not comprise any ligand information, the question if these compounds are similar to what we would expect of D2R ligands arises. To provide an answer, compounds with a score higher than 0.5 generated from runs with standard settings (*i.e.*, with one augmentation and a learning rate of 0.001), with a moderately increased number of augmentations (*i.e.*, five) or with an increased learning rate (*i.e.*, 0.005) were compared to known D2R ligands by calculating their pairwise Tanimoto Morgan fingerprints similarity. As a baseline, the dataset produced using the standard settings had two compounds with a similarity higher than 0.8 and 84 compounds with a similarity higher than 0.7. Increasing the number of augmentations in the inner loop to five increased the number of identified compounds with a similarity higher than 0.8 to 24 and the number of compounds with a similarity higher than 0.7 to 140. In contrast, increasing the learning rate led to a drop in the number of compounds with a similarity higher than 0.7 to 41 but did not affect the number of compounds with a similarity higher than 0.8 with two such compounds produced.

The ten compounds of highest similarity to known D2R binders from the run with five augmentations are shown alongside their nearest neighboring known D2R binders in Figure 11. Given the wide-ranging generative possibilities and the lack of ligand structure guidance for the generation of ligands, as would be available from a scoring function comprising a QSAR model, it is noteworthy that it is possible to generate compounds with a relatively high structural similarity to known ligands by only using a docking target. Nevertheless, the obtained molecules are typically small and do not achieve particularly high docking scores, which can likely be attributed to the ease of generating a molecule with a high similarity to another when both molecules are relatively small. Many of the generated compounds (and known ligands) share chemical features with the native D2R receptors ligand, *i.e.*, dopamine, but also comprise a locked aliphatic amine. Overall, the marked increase in the number of high similarity compounds generated by the run with five augmentations indicates that the double loop augmented reinforcement learning increases the ability of the algorithm to find similar compounds to the native ligand and other known binders.

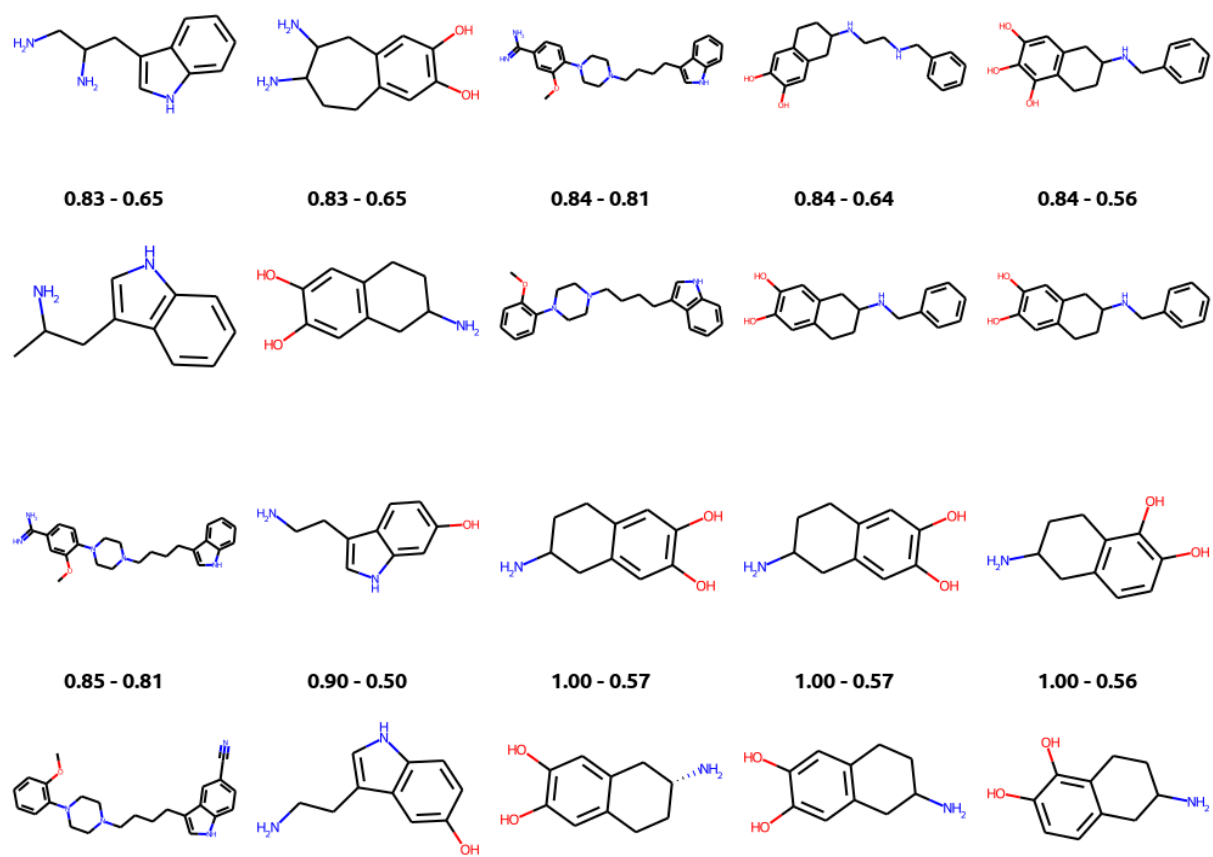

*Figure 11: Generated compounds with highest similarity to known D2R binders. The generated compounds are shown in rows 1 and 3, and their closest known ligand neighbors are shown in rows 2 and 4. The numbers denote the Tanimoto similarity between the pairs and the score obtained for the generated compound.*

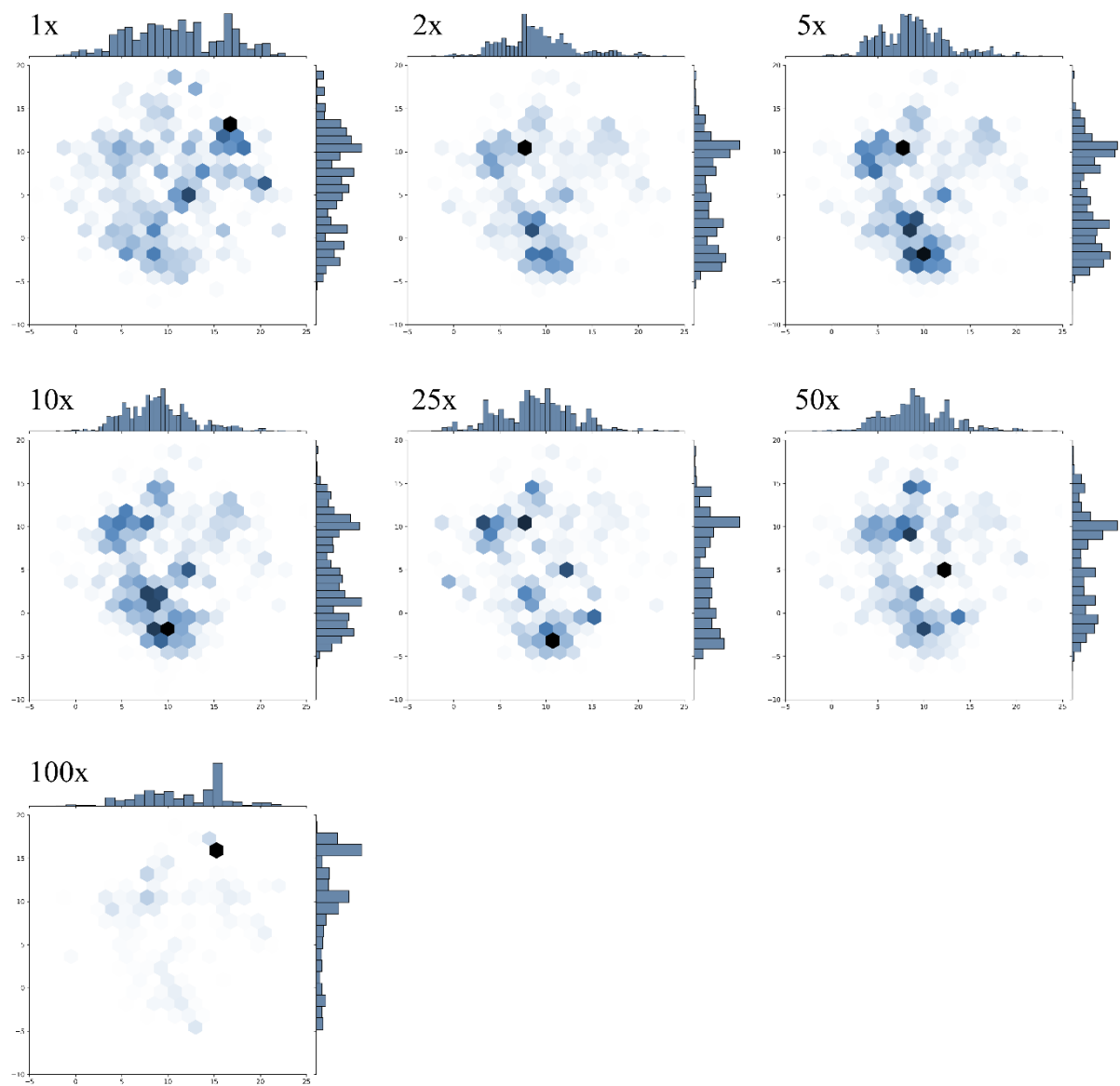

*Figure 12 ChemCharts hexbin plot visualization of the distribution of sampled molecules from pooled triplicate runs of the docking task with different levels of augmentation. Augmentations increase left-to-right, top-to-bottom. Runs employing between 2 and 10 augmentations sample very similar spaces with a high diversity.*

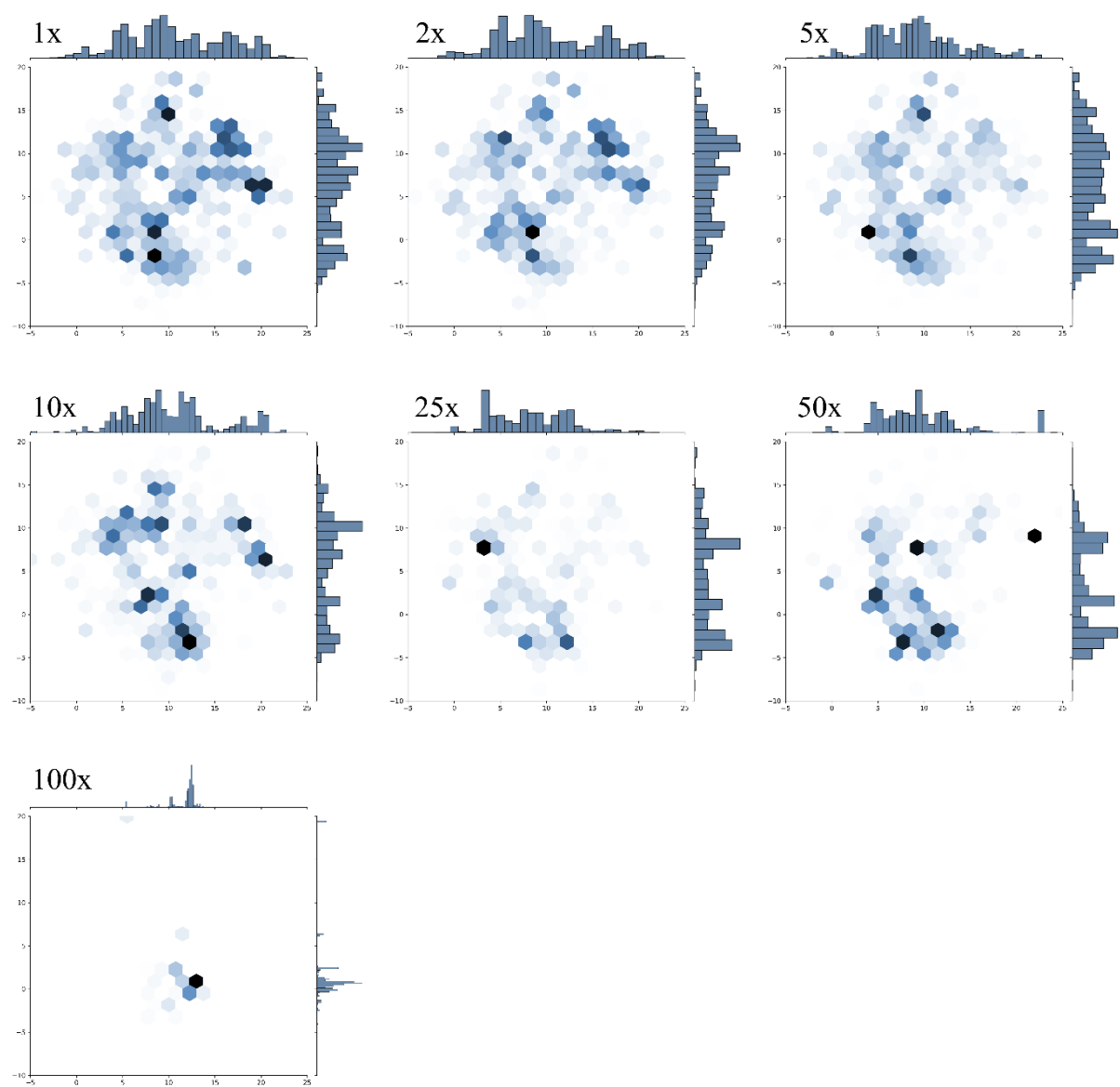

*Figure 13 ChemCharts hexbin plot visualization of the distribution of sampled molecules from pooled triplicate runs of the docking task with different learning rates. Learning rate multiplication factors increase row wise from 1 (top-left) to 100 (bottom). Runs with a learning rate multiplication factor of between 1 and 5 seem to sample the same chemical areas but with different intensity. As the learning rate increases fewer areas seem to be sampled until full mode collapse is observed at a learning rate multiplication factor of 100.*

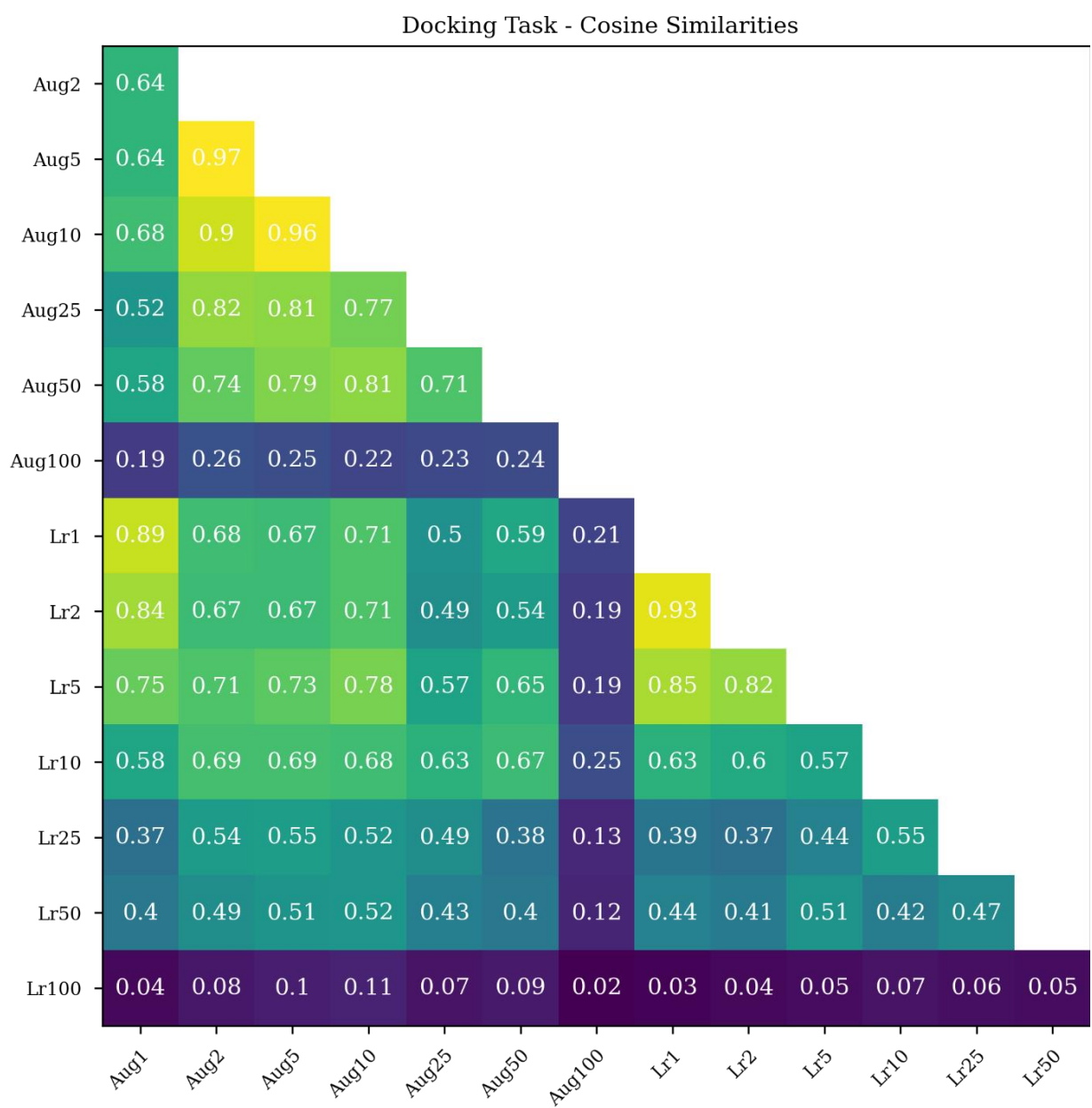

Figure 14: ChemCharts cosine similarities between the pooled triplicate generative runs. Light yellow color indicates a large similarity between the sampled chemical spaces, whereas dark purple indicates a low similarity.

## Assessment of augmentation effect on standard and hill-climb reinforcement learning

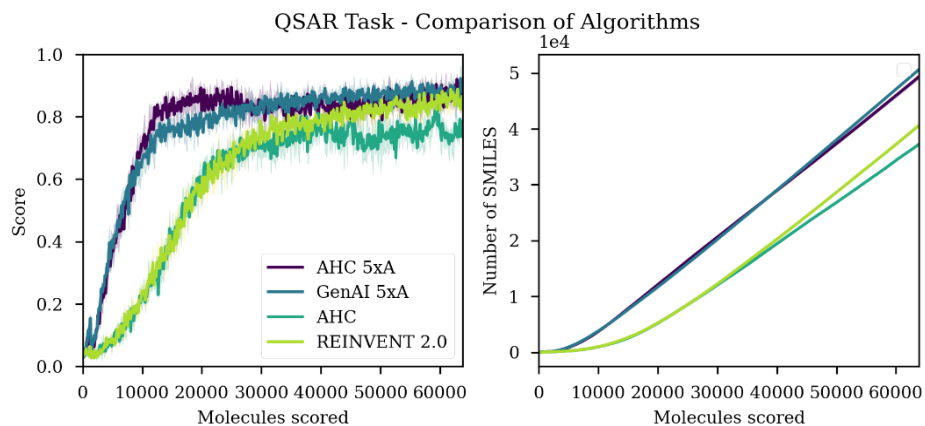

*Figure 15: Effect of algorithm type on generative performance. LEFT: Average score of generated molecules as a function of the number of scored molecules. RIGHT: Number of generated molecules with an average score above 0.5 as a function of the number of scored molecules. All experiments were performed in triplicate.*

To assess the improvement in generative performance brought about by the use of SMILES augmentation, the optimization towards a D2R QSAR classification model of four generative algorithms was compared: REINVENT 2.0; *GenAI*, *i.e.*, REINVENT 2.0 with an inner SMILES augmentation loop (here, with five augmentations); augmented hill-climb (*AHC*)[14]; and AHC with an inner SMILES augmentation loop (*AHC 5xA*) also with five augmentations.

First, the influence of introducing an inner loop of five SMILES augmentations on the reinforcement learning efficiency and the generated compounds yield was assessed. The plot in Figure 15 (left panel) suggests that the algorithms employing SMILES augmentation, *i.e.*, *GenAI* and *AHC 5xA*, both reached higher average scores faster and converged markedly quicker than the algorithms that did not use SMILES augmentation, *i.e.*, REINVENT 2.0 and *AHC*. Interestingly, while the *AHC 5xA* model reached average scores above 0.8 first, *GenAI* typically achieved the highest average scores among the algorithms after 30,000 molecules were scored. In contrast, the standard *AHC* algorithm performed slightly worse than REINVENT 2.0 with minor differences in efficiency being accompanied by a convergence to notably lower scoring values. When considering the number of unique generated SMILES with a score higher than 0.5 as a function of the number of scored molecules plotted in Figure 15 (right panel), the differences between the four generative strategies closely mirror the reinforcement learning efficiency trends. Specifically, the algorithms utilizing SMILES augmentation both yielded a much higher number of unique well-scoring SMILES than their non-augmented counterparts and, again, *GenAI* generated the largest number of SMILES, while the standard AHC model produced the lowest yields. Additionally, while the AHC algorithm become productive after approximately the same number of molecules were scored as REINVENT 2.0, its rate of generating new molecules was lower. Overall, the augmented strategies enabled the generation of higher scoring molecules faster and were able to produce a larger number of unique well-scoring molecules than their non-augmented counterparts. Further, the average yield of the different models over all experimental repeats as a function of the threshold score was plotted in Figure 16. The augmented strategies were observed to perform better than the non-augmented ones. This agrees with the results above, which also indicate that the introduction of an inner SMILES augmentation loop is correlated with higher average scores across the full duration of the generative runs. Interestingly, the AHC-based algorithms, *i.e.*, *AHC* and *AHC 5xA*, achieved a higher fraction of top-scoring molecules than their non-AHC-based counterparts, *i.e.*, REINVENT 2.0 and *GenAI*, respectively, as evidenced by the cross-over and steepness of the yield curves close to score values of one.

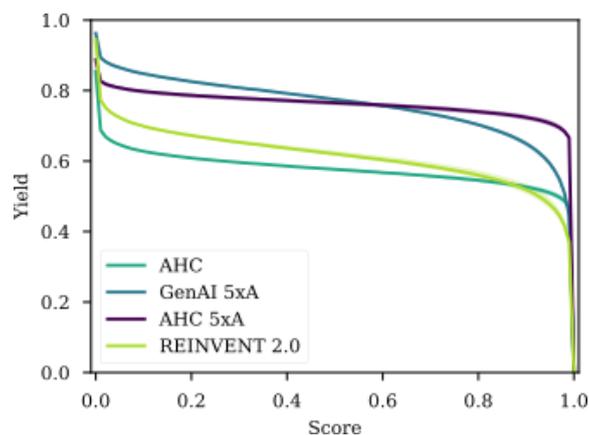

*Figure 16: Yield fraction curves as a function of the score. Higher yield represents the generation of larger numbers of unique SMILES with a higher score.*

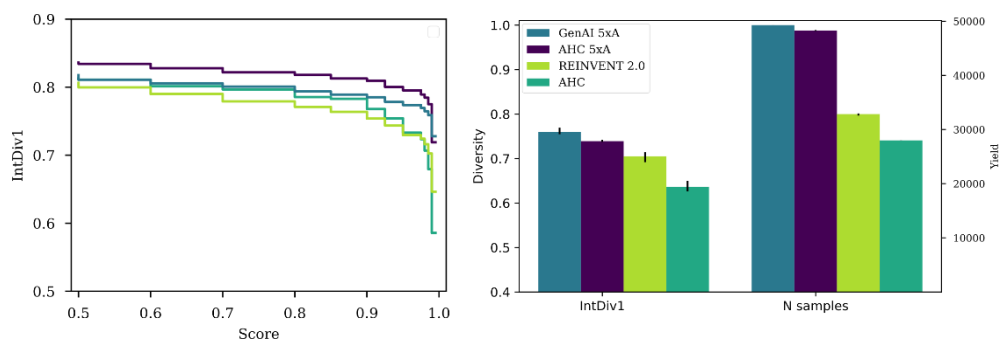

*Figure 17: Internal diversity properties of the datasets produced using the four generation algorithms. LEFT: Intervallic internal diversity (IntDiv1) as a function of the score thresholds. RIGHT: Overall internal diversity of unique molecules with a score greater than 0.5 and number of unique molecules generated. The error-bars show the min and the max obtained from the triplicate runs.*

Next, the internal diversity of the compounds generated using the four algorithms of interest was examined. The total internal diversity of unique molecules with a score greater than 0.5 is shown in Figure 17 (right panel) with the average values for the triplicate runs being: 0.76 for *GenAI*, 0.74 for *AHC 5xA*, 0.70 for REINVENT 2.0 and 0.63 for the *AHC* algorithm. These results suggest that the augmented strategies not only reached higher scores faster (see above) but the molecules produced with them are also considerably more diverse with *GenAI* giving the highest number of diverse molecules. Additionally, the average internal diversity across different score thresholds was analyzed in Figure 17 (left panel) with, for example, the first interval in the graph representing the internal diversity of unique generated molecules that have a score between 0.5 and 0.6. Given that the augmented algorithms were shown to generate molecular datasets with a higher total internal diversity and a higher yield (see above), it is not surprising that the same performed better here too. It is worth noting that despite the *AHC* and *AHC 5xA* strategies achieving higher internal diversities over most thresholds in comparison to REINVENT 2.0 and *GenAI,* respectively, when the threshold was set close to the maximum score of one the non-AHC strategies achieved higher internal diversities that their AHC-based counterparts. In other words, the AHC-based strategies produced less diverse compounds with scores close to one, which is also the scores range in which AHC algorithms obtained a much higher yield than their non-AHC counterparts (*cf.* Figure 16). This, in turn, can explain the fact that the total diversity for AHC models is lower than that for their non-AHC counterparts when all molecules with a score above 0.5 are considered (*cf.* Figure 17 (right panel)). Together, the results indicate that algorithms implementing SMILES augmentation produce molecules of a greater overall diversity, as well as more diverse high scoring compounds.

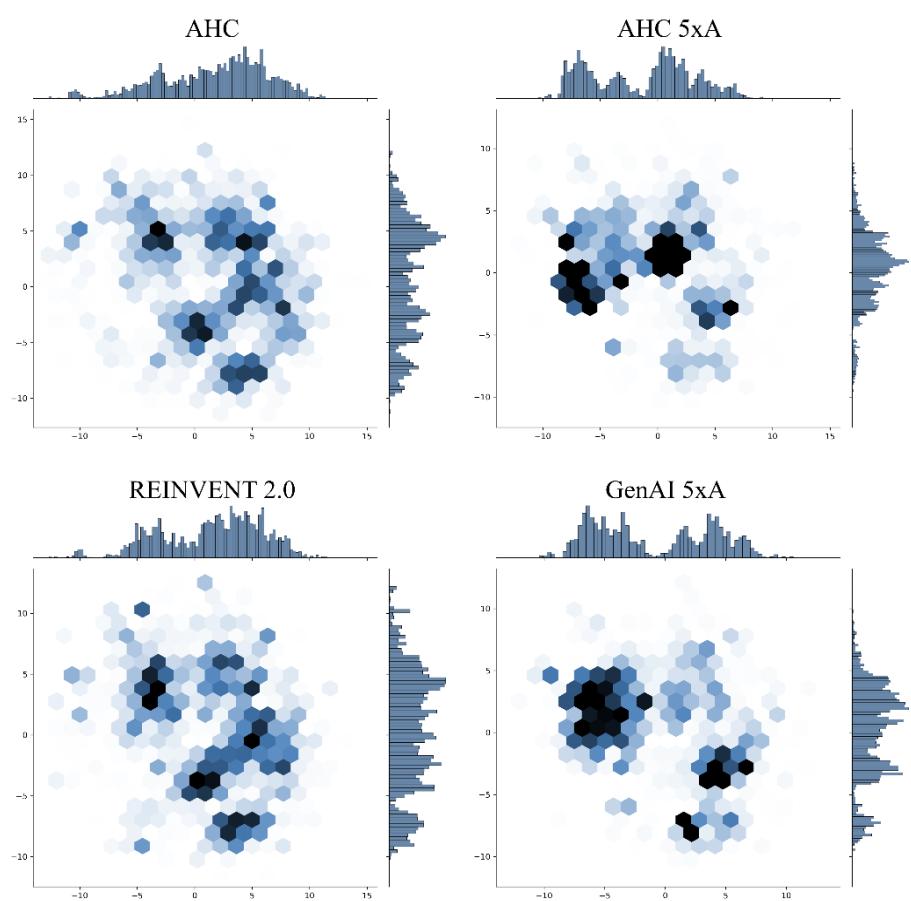

*Figure 18: Normalized ChemCharts with hex visualization of the distribution in chemical space of the molecules generated by the different algorithms. The two non-augmented models appear visually similar. The augmented algorithms appear to sample fewer areas but focus on neighboring hex-bins more intensely.*

The chemical space covered by the molecules that were generated by each of the four generative algorithms and had a score higher than 0.5 in all three experimental repeats is shown in the 2D hexbin ChemCharts plots in Figure 18. All datasets encompass similar areas within the chemical space; however, the molecules produced by the non-augmented strategies covered a wider chemical region less densely. A possible explanation for these observations is that the augmented strategies were more likely to focus on narrow chemical areas since they generated higher scoring molecules earlier in the generative runs (*cf.* Figure 15), while the non-augmented models were instead more prone to exploring a wider range of areas in search of higher scoring molecules (*cf.* Figure 15). Despite the non-augmented strategies datasets covering a slightly wider area of the chemical space, the internal diversity plots show that the augmented strategies generated datasets of higher diversity (*cf.* Figure 17). Interestingly, the non-augmented runs are more similar than to their respective augmented counterpart. This trend was also observed for the docking runs, where lr1 is more similar to aug1, than they are to lr2 and aug2, respectively (*cf.* Figure 14). Specifically, the two non-augmented strategies, *i.e.*, REINVENT 2.0 and *AHC*, covered the most similar areas of chemical space with a ChemCharts cosine similarity value of 0.79, *GenAI* produced compounds with a similarity to the rest of the datasets ranging between 0.44 and 0.48, and the augmented AHC model appeared to produce the most dissimilar SMILES to those generated with non-augmented algorithms with cosine similarities of 0.27-0.28.

Overall, it is clear from the discussed results shown in Figure 14-16 that *GenAI 5xA* outperformed both the non-augmented generative algorithms evaluated here, *i.e.*, REINVENT 2.0 and *AHC*, as well as the *AHC 5xA* algorithm in terms of overall SMILES yield and diversity, with the exception of the yield obtained for perfectly scoring molecules, for which only the *AHC 5xA* algorithm performed better. The AHC strategy seems to lead to an increase in the fraction of perfectly scoring molecules but with a decrease in diversity when compared to the standard reinforcement implemented in REINVENT 2.0; however, the effect on yield and diversity can be largely counteracted by introducing SMILES augmentations as is done in *AHC 5xA*.

## Discussion

The results presented clearly demonstrate that adding multiple augmentations in the inner loop has a different impact on *de novo* molecular optimization compared to increasing the learning rate. Not only does adding more augmentations result in generative runs that can more consistently achieve higher scores faster and are more robust against instabilities at high hyperparameter settings, but it also improves the quality of the generated SMILES dataset to obtain higher yields, higher internal diversity, and more generated molecules highly similar to known ligands. This may be because the use of augmentations encourages the algorithm to focus more on the "molecules" rather than the specific sequences generated, in the sense that the agent remains in the augmented SMILES space. Similarly, SMILES augmentation has been shown to improve the QSAR feature relevance of the latent space generated in autoencoder training[22], which can be interpreted as increased molecular relevance. In fact, the prior network used here has previously been trained using SMILES augmentation, which was shown to improve chemical space coverage, validity, and in-scope generation in unconditional generation[16].; hence, extending SMILES augmentation to the reinforcement learning is likely to be able to bring similar benefits. When using the standard generative algorithm setting, reinforcement learning encourages the agent to generate similar sequences, which can lead to lower SMILES diversity and, ultimately, mode-collapse. In contrast, when employing the proposed augmented setting, the agent is never reinforced on the sequences it has generated, as these are augmented and thus changed before being used to reinforce the agent. This effectively prevents the agent from over-focusing on specific SMILES sequences, thereby leading to lower propensity for sequence-wise mode-collapse and higher uniqueness and internal diversity as a result.

While recent publications[14], [19] discussing the optimization efficiency of different molecular generative algorithms have primarily focused on the speed with which such models can produce well-scoring compounds and hence be exploited, less emphasis has been placed on the diversity and yield of the generated compounds. This could perhaps be attributed to the ease with which exploitation efficiency can be quantified by a single metric, whereas the evaluation of the diversity and quality of generated molecules is often not as straight-forward. As a result, many articles published over the last years have showcased and, indeed, competed to develop narrowly focused algorithms which can most efficiently find the compound with, for instance, the largest calculated penalized logP.[37] However, this is often not a useful example scenario in drug discovery as the aim of generation is seldom to maximize a single property but rather more commonly to find the balance between a range of often conflicting molecular properties to obtain a suitable drug candidate.[38] Indeed, most often generative *de novo* design in drug discovery is performed with an aspect of exploration and diversity in mind and with the aim of producing several putative ligand series to serve as ideas for continuous development through filtering and prioritization based on criteria such as synthesizability, and in collaboration with computational chemists and medicinal chemist. Hence, the diversity of the generated compounds needs to be considered for a comprehensive view of the performance of a generative algorithm to be obtained. As our results suggest, varying a single tuning parameter, such as the learning rate, can increase the efficiency of generation, however, this may be accompanied by a tradeoff in the diversity and yields of the produced compounds. The introduction of an inner SMILES augmentation loop, on the other hand, allows for both an increase in the efficiency as well the diversity and reproducibility of *de novo* generation, thereby making the strategy particularly attractive for use in practical drug discovery scenarios.

To assess the performance of the proposed augmented generative algorithm, *i.e.*, *GenAI*, the increase in the efficiency speed of generation was considered first. In a recent study, the augmented hill-climbing (AHC) algorithm showed an impressive 45-fold speed-up using sample efficiency for first sampled compound and a 7.4-fold increase in efficiency on average compared to the baseline REINVENT 1.0.[14] The sample efficiency is the generative step, for which the score achieved for a single generated compound or the average score achieved for all generated compounds surpasses a given threshold. Although we have not analyzed the single sample efficiency in a similar manner in this study, we can estimate the increase in speed with which the augmented *GenAI* models can produce well-scoring compounds compared to the standard REINVENT 2.0 algorithm using the data shown in Table 1.

Specifically, our results suggest that for the average scoring thresholds of 0.5, 0.8 and single compound solved, an increase in efficiency of ~10, ~4 and ~2 over the baseline can be observed when the number of inner loop SMILES augmentations is set to 100, 20 or 10, respectively. Hence, the efficiency speed-up appears to be strongly dependent on the scoring threshold and solution criteria used. While the increases in efficiency reported here are comparably smaller than those reported by *Thomas et al.* [14], it should be noted that the literature study compared the proposed AHC algorithm to the first version of REINVENT, which struggled to obtain meaningful docking scores for the D2R docking target within 500 steps. In contrast, REINVENT 2.0 was used as a baseline in this study and showed marked improvements in the docking scores of compounds after only ~7500 molecules (corresponding to ~120 steps) were scored. Additionally, the sigma parameter of the REINVENT algorithm was also altered in the AHC algorithm preprint[14], whereas in this study the default value of 128 was used. Hence, it can be deduced that the baseline algorithms used to determine the increases in efficiency in the two studies are not identical or performing comparably and, hence, the reported improvements can also not be directly compared. Generally, we find that using a sample-level efficiency increase as a quantification for the performance of generative models is often uninformative, since the metric relies on the selection of a baseline model, as well as the studied algorithms finding a single well-scoring molecule and does not account for average scores or the diversity of the generated molecules. Nevertheless, the proposed double loop augmented reinforcement learning algorithm seems to be competitive when assessed using the average score efficiency speed-up. Thomas and co-workers also observed that in the absence of a diversity filter the AHC algorithms were prone to mode collapse. Hence, in this study, we employed a standard diversity filter for the AHC models; nevertheless, we still observed a tendency for the AHC models to yield a relatively high fraction of perfectly scoring molecules, but of comparably low internal diversity, which could be an initial indication for hyper-focus and the beginning of mode collapse.

Next, we assessed the proposed generative algorithm augmentation in the context of an extensive generative benchmark recently developed by *Gao et al.*[19], which primarily focuses on optimization efficiency. In the study, the authors benchmarked 25 generative methods across 23 widely used oracle functions. The performance of the different algorithms and models was compared after 10,000 queries to the scoring function, using the area under the scoring curve (AUC) for the top 10 solutions found.[19] Based on this model comparison, the authors reported that REINVENT 2.0 performs competitively among the selection of algorithms, albeit with a tuned sigma value.[19] These results support our choice of REINVENT 2.0 as a high-quality baseline and starting point for this work. To compare the performance of the augmented algorithms, *i.e.*, *GenAI* and augmented AHC, to that of the benchmarked generative algorithms, we selected the benchmarking task which most closely resembled the QSAR task in our study, namely, the DRD2 oracle - a QSAR model of a D2R dataset. We calculated the AUC for the top 10 highest scoring compounds in our QSAR task in the same manner as in the benchmark by Gao et al.[19] The resulting AUC values for *GenAI* and the augmented AHC were 0.9989 ± 0.0003 and 0.9985 ± 0.0004, respectively. However, we believe that these high values may be attributed to differences in the QSAR models employed given that the standard REINVENT 2.0 run also achieved a high value of 0.9977±0.0004, which is markedly higher than the 0.945±0.007 obtained for it in the PMO benchmark.[19] Our QSAR model is not calibrated with regard to probability and has a tendency to produce values close to either zero or one, which could explain the differences. As briefly mentioned above, to benchmark the performance of REINVENT 2.0 Gao and co-workers tuned the algorithm's sigma parameter.[19] By increasing the value of sigma the authors likely improved the model's efficiency; however, this change could also have negatively affected the diversity and yield on the generated compounds and, moreover, since the weight of the prior generator is decreased, the quality of the proposed molecules could have also suffered.[20] In contrast, the performance of ReInvent 1.0 was observed to be relatively unaffected by changes in sigma.[14] In this study, the sigma variable for reinforcement learning was fixed at the default value of 128, which is an aspect that could be investigated more in the future.

Another difference between the settings used in the discussed benchmark and our work is the length of the generative experiment. Specifically, Gao and co-workers compared the performance of models over 10,000 oracle calls, which corresponds to approximately 150 generative steps with a batch size of 64, whereas experiments in this study were conducted for between 500 and 1000 steps of the same batch size. As shown by the results of several of our experiments, the selection of best performing algorithm and/or parameters can vary with the length of the generative run. For instance, if experiments of 150 steps were used to select the best learning rate for the docking task, the conclusion that a learning rate multiplication factor of 5 leads to the greatest efficiency would have been reached, however, running the experiments for longer, *i.e.*, 500 steps, allowed for the decrease in SMILES yield with increase in learning rate to be observed and, thus, change the conclusion regarding which hyperparameter values lead to the most efficient algorithm (*cf.* Figure 7). Additionally, 10,000 function evaluations are an order of magnitude lower than what we would employ in generative runs in practice for docking scores. Hence, running the benchmarks for longer and reporting the findings for different experimental run lengths could be beneficial for model selection in practical applications and would allow for the computational budget of oracle calls to be taken into consideration when designing generative experiments.

Lastly, when comparing the varying algorithms, Gao and co-workers consider solely the ten highest scoring compounds, which, in our experience, is a far too small number of molecules to progress to further post-processing. Moreover, the resulting benchmarking metrics are narrowly focused on quantifying how well the benchmarked algorithms can generate a few well-scoring candidates, which, in turn, leads to the majority of the attention being placed on optimization and less consideration for the diversity and yield of the methods. Indeed, the authors discuss this drawback in their work and note that they plan to address it in a future update of the PMO benchmark.

Overall, we were pleased to report that the proposed augmented algorithm can generate ligands similar to known ligands even in the absence of a ligand-based scoring function but solely utilizing a docking target. Despite these encouraging results, it must be noted that the prior used in our work and, hence, also the agent are pre-trained on a large chemical dataset that likely includes SMILES strings from D2R datasets used here. This could result in the prior having a propensity for generating compounds with a structure similar to that of the ligands in the D2R dataset; however, the fraction of D2R ligands in the prior is likely very small and their effect is probably negligible. Indeed, it has previously been shown that removing known or similar ligands associated with a target does not hinder reinforcement learning from rediscovering known ligands.[2], [39]

Recently, the barrier to obtaining structural information for protein targets has been greatly diminished with the introduction of algorithms such as AlphaFold2, which can produce high-quality 3D protein structural prediction.[40] Thus, the development of a robust *in silico* pathway from target sequence to ligand seem within reach, albeit that using predicted protein structures for docking is not yet as reliable as utilizing X-ray crystallographic data.[41]

# Conclusion

The experimental results described in this work suggest that the efficiency of reinforcement learning for optimization of molecular generation can be increased by both increasing the learning rate of REINVENT 2.0 (*i.e.*, allowing for bigger weight updates to the agent per step) or by increasing the number of augmentations as is done in *GenAI* (*i.e.*, allowing multiple smaller weight updates to the agent per step to reinforce different generation paths to the same molecules). However, increasing the efficiency can also be associated with a lower yield and/or diversity of the generated compounds. Nevertheless, using a moderate number of augmentations (*i.e.*, five or ten) can both improve the reinforcement learning efficiency, as well as increase the diversity of the produced SMILES; whereas increasing the learning rate negatively affects the final yield even for the smallest multiplication factor investigated here, *i.e.*, two. Furthermore, increasing the number of inner loop SMILES augmentation to between two and ten was observed to improve the reproducibility of the results in terms of the chemical space sampled as well as generating more compounds closely similar to known ligands using only a docking score as target. Our recommendation is to use around five inner augmentation loops for exploratory runs and up to ten augmentations for more exploitative scenarios. These settings have shown to produce more efficient and diverse generative runs in our experiments.

# Acknowledgements

We want to acknowledge the data science team at Odyssey Therapeutics for their helpful feedback and discussions, and, especially, Dr. Atanas Patronov and Dr. Kostas Papadopoulus for their REINVENT 2.0 expertise. We also want to thank Sophie Margreitter for the helpful discussions regarding ChemChart code modifications.

# Conflicts of Interest

Authors are employees at Odyssey Therapeutics, which has a commercial interest in utilizing generative modelling of prospective drug candidates.

# Code availability

Code was implemented in the proprietary codebase, *GenAI*, of the Odyssey Therapeutics generative drug discovery platform.

# Contributions

Dr. Esben Jannik Bjerrum made the initial approach suggestion, implemented the prototype and production code for *GenAI*, carried out and analyzed the similarity and docking tasks, wrote the draft manuscript and had overall supervision of the project. Raquel Lopez-Rios de Castro ported the augmented Hill-Climb code modifications to *GenAI*, developed the D2R QSAR model, and carried out and analyzed the QSAR task and the comparison of the AHC algorithm variations. Dr. Christian Margreitter and Dr. Thomas Blaschke provided help with the setup of the REINVENT 2.0 algorithm and scoring functions and offered helpful discussions and feedback on results. Dr. Simona Kolarova offered helpful discussions and feedback, and extensively helped with the writing of the manuscript. All authors read, edited, and approved the final paper.

# References

[1] E. J. Bjerrum and R. Threlfall, "Molecular Generation with Recurrent Neural Networks (RNNs)." May 2017. [Online]. Available: http://arxiv.org/abs/1705.04612

[2] M. Olivecrona, T. Blaschke, O. Engkvist, and H. Chen, "Molecular de-novo design through deep reinforcement learning," *J. Cheminformatics*, vol. 9, no. 1, p. 48, Sep. 2017, doi: 10.1186/s13321-017-0235-x.

[3] M. H. S. Segler, T. Kogej, C. Tyrchan, and M. P. Waller, "Generating focused molecule libraries for drug discovery with recurrent neural networks," *ACS Cent. Sci.*, vol. 4, no. 1, pp. 120–131, Jan. 2018, doi: 10.1021/acscentsci.7b00512.

[4] A. Kadurin, S. Nikolenko, K. Khrabrov, A. Aliper, and A. Zhavoronkov, "DruGAN: An Advanced Generative Adversarial Autoencoder Model for de Novo Generation of New Molecules with Desired Molecular Properties in Silico," *Mol. Pharm.*, vol. 14, no. 9, pp. 3098–3104, Sep. 2017, doi: 10.1021/acs.molpharmaceut.7b00346.

[5] R. Gómez-Bombarelli *et al.*, "Automatic Chemical Design Using a Data-Driven Continuous Representation of Molecules," *ACS Cent. Sci.*, vol. 4, no. 2, pp. 268–276, Feb. 2018, doi: 10.1021/acscentsci.7b00572.

[6] Bjerrum, Esben Jannik, "Teaching Computers Molecular Creativity," *Cheminformania*, Nov. 2016. https://www.cheminformania.com/teaching-computers-molecular-creativity/

[7] M. H. S. Segler, T. Kogej, C. Tyrchan, and M. P. Waller, "Generating Focussed Molecule Libraries for Drug Discovery with Recurrent Neural Networks." arXiv, Jan. 05, 2017. Accessed: Feb. 06, 2023. [Online]. Available: http://arxiv.org/abs/1701.01329

[8] T. Blaschke *et al.*, "REINVENT 2.0: An AI Tool for De Novo Drug Design," *J. Chem. Inf. Model.*, vol. 60, no. 12, pp. 5918–5922, Dec. 2020, doi: 10.1021/acs.jcim.0c00915.

[9] J. Guo *et al.*, "Link-INVENT: Generative Linker Design with Reinforcement Learning." Apr. 25, 2022. doi: 10.26434/chemrxiv-2022-qkx9f.

[10] V. Fialková *et al.*, "LibINVENT: Reaction-based Generative Scaffold Decoration for in Silico Library Design," *J. Chem. Inf. Model.*, p. acs.jcim.1c00469, Aug. 2021, doi: 10.1021/acs.jcim.1c00469.

[11] D. C. Elton, Z. Boukouvalas, M. D. Fuge, and P. W. Chung, "Deep learning for molecular design - A review of the state of the art," *Mol. Syst. Des. Eng.*, vol. 4, no. 4, pp. 828–849, 2019, doi: 10.1039/c9me00039a.

[12] Y. Xu *et al.*, "Deep learning for molecular generation," *Future Med. Chem.*, p. fmc-2018-0358, Jan. 2019, doi: 10.4155/fmc-2018-0358.

[13] M. Wang *et al.*, "Deep learning approaches for de novo drug design: An overview," *Curr. Opin. Struct. Biol.*, vol. 72, pp. 135–144, Feb. 2022, doi: 10.1016/j.sbi.2021.10.001.

[14] M. Thomas, N. M. O'Boyle, A. Bender, and C. de Graaf, "Augmented Hill-Climb increases reinforcement learning efficiency for language-based de novo molecule generation." Apr. 15, 2022. doi: 10.26434/chemrxiv-2022-prz2r.

[15] E. J. Bjerrum, "SMILES Enumeration as Data Augmentation for Neural Network Modeling of Molecules." arXiv, May 17, 2017. doi: 10.48550/arXiv.1703.07076.

[16] J. Arús-Pous *et al.*, "Randomized SMILES strings improve the quality of molecular generative models," *J. Cheminformatics*, vol. 11, no. 1, pp. 1–13, 2019.

[17] D. Neil *et al.*, "Exploring deep recurrent models with reinforcement learning for molecule design," *6th Int. Conf. Learn. Represent. ICLR 2018 - Workshop Track Proc.*, pp. 1–15, 2018.

[18] N. Brown, M. Fiscato, M. H. S. Segler, and A. C. Vaucher, "GuacaMol: Benchmarking Models for de Novo Molecular Design," *J. Chem. Inf. Model.*, vol. 59, no. 3, pp. 1096–1108, Mar. 2019, doi: 10.1021/acs.jcim.8b00839.

[19] W. Gao, T. Fu, J. Sun, and C. W. Coley, "Sample Efficiency Matters: A Benchmark for Practical Molecular Optimization." arXiv, Jun. 22, 2022. doi: 10.48550/arXiv.2206.12411.

[20] P. Renz, D. Van Rompaey, J. K. Wegner, S. Hochreiter, and G. Klambauer, "On failure modes in molecule generation and optimization," *Drug Discov. Today Technol.*, vol. 32–33, pp. 55–63, 2019, doi: 10.1016/j.ddtec.2020.09.003.

[21] P.-C. Kotsias, J. Arús-Pous, H. Chen, O. Engkvist, C. Tyrchan, and E. J. Bjerrum, "Direct steering of de novo molecular generation with descriptor conditional recurrent neural networks," *Nat. Mach. Intell.*, vol. 2, no. 5, pp. 254–265, 2020.

[22] E. J. Bjerrum and B. Sattarov, "Improving chemical autoencoder latent space and molecular de novo generation diversity with heteroencoders," *Biomolecules*, vol. 8, no. 4, p. 131, 2018.

[23] R. Irwin, S. Dimitriadis, J. He, and E. J. Bjerrum, "Chemformer: a pre-trained transformer for computational chemistry," *Mach. Learn. Sci. Technol.*, vol. 3, no. 1, p. 015022, Jan. 2022, doi: 10.1088/2632-2153/ac3ffb.

[24] D. Sumner, J. He, A. Thakkar, O. Engkvist, and E. J. Bjerrum, "Levenshtein Augmentation Improves Performance of SMILES Based Deep-Learning Synthesis Prediction." 2020. doi: 10.26434/chemrxiv.12562121.v1.

[25] D. Polykovskiy *et al.*, "Molecular Sets (MOSES): A Benchmarking Platform for Molecular Generation Models," *Front. Pharmacol.*, vol. 11, 2020, Accessed: Aug. 25, 2022. [Online]. Available: https://www.frontiersin.org/articles/10.3389/fphar.2020.565644

[26] S. Margreitter, "ChemCharts." Sep. 07, 2022. Accessed: Sep. 09, 2022. [Online]. Available: https://github.com/SMargreitter/ChemCharts

[27] T. Blaschke, O. Engkvist, J. Bajorath, and H. Chen, "Memory-assisted reinforcement learning for diverse molecular de novo design," *J. Cheminformatics*, vol. 12, no. 1, pp. 1–17, 2020, doi: 10.1186/s13321-020-00473-0.

[28] "ReinventCommunity (jupyter notebook tutorials for REINVENT 3.2)." AstraZeneca - Molecular AI, https://github.com/MolecularAI/ReinventCommunity, Sep. 08, 2022. Accessed: Sep. 09, 2022. [Online]. Available: https://github.com/MolecularAI/ReinventCommunity

[29] S. Wang, T. Che, A. Levit, B. K. Shoichet, D. Wacker, and B. L. Roth, "Structure of the D2 dopamine receptor bound to the atypical antipsychotic drug risperidone," *Nature*, vol. 555, no. 7695, pp. 269–273, Mar. 2018, doi: 10.1038/nature25758.

[30] H. M. Berman *et al.*, "The Protein Data Bank," *Nucleic Acids Res.*, vol. 28, no. 1, pp. 235–242, Jan. 2000, doi: 10.1093/nar/28.1.235.

[31] R. A. Friesner *et al.*, "Glide:  A New Approach for Rapid, Accurate Docking and Scoring. 1. Method and Assessment of Docking Accuracy," *J. Med. Chem.*, vol. 47, no. 7, pp. 1739–1749, Mar. 2004, doi: 10.1021/jm0306430.

[32] J. Guo *et al.*, "DockStream: a docking wrapper to enhance de novo molecular design," *J. Cheminformatics*, vol. 13, no. 1, p. 89, Nov. 2021, doi: 10.1186/s13321-021-00563-7.

[33] J. Sun *et al.*, "ExCAPE-DB: an integrated large scale dataset facilitating Big Data analysis in chemogenomics," *J. Cheminformatics*, vol. 9, no. 1, p. 17, Mar. 2017, doi: 10.1186/s13321-017-0203-5.

[34] F. Pedregosa *et al.*, "Scikit-learn: Machine Learning in Python," *J. Mach. Learn. Res.*, vol. 12, no. null, pp. 2825–2830, Nov. 2011.

[35] "RDKIT: Open source cheminformatics." [Online]. Available: http://www.rdkit.org

[36]  L. McInnes, J. Healy, and J. Melville, "UMAP: Uniform Manifold Approximation and Projection for Dimension Reduction." arXiv, Sep. 17, 2020. doi: 10.48550/arXiv.1802.03426.

[37]  R. Richards and A. Groener, "Conditional $\beta$-VAE for De Novo Molecular Generation," May 2022, doi: 10.26434/chemrxiv-2022-g3gvz.

[38]  L. Di and E. H. Kerns, *Drug-Like Properties: Concepts, Structure Design and Methods from ADME to Toxicity Optimization*, 2nd ed. Elsevier, 2016.

[39]  K. Papadopoulos, K. A. Giblin, J. P. Janet, A. Patronov, and O. Engkvist, "De novo design with deep generative models based on 3D similarity scoring," *Bioorg. Med. Chem.*, vol. 44, p. 116308, Aug. 2021, doi: 10.1016/j.bmc.2021.116308.

[40]  J. Jumper *et al.*, "Highly accurate protein structure prediction with AlphaFold," *Nature*, vol. 596, no. 7873, Art. no. 7873, Aug. 2021, doi: 10.1038/s41586-021-03819-2.

[41]  F. Wong *et al.*, "Benchmarking AlphaFold-enabled molecular docking predictions for antibiotic discovery," *Mol. Syst. Biol.*, vol. 18, no. 9, p. e11081, Sep. 2022, doi: 10.15252/msb.202211081.

# Supplementary Material for "Faster and more diverse *de novo* molecular generation with double-loop reinforcement learning optimization using augmented SMILES"

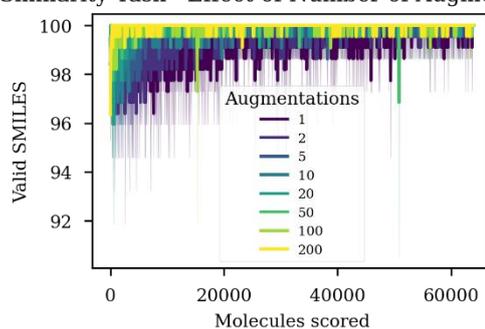

*Supplementary Figure 1.* Effect of the number of augmentations on the percentage of valid SMILES from all generated SMILES obtained from the similarity task. All generative runs produced sets of SMILES with percentage validity in the high nineties, which appeared to slightly increase with the number of augmentations.

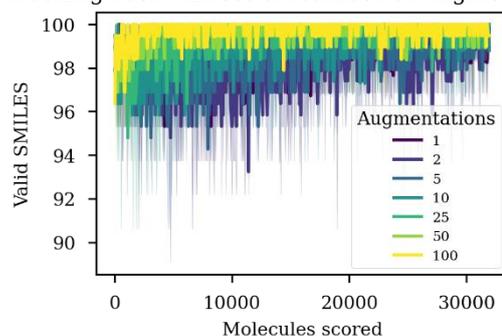

*Supplementary Figure 2.* Effect of the number of augmentations on the percentage of valid SMILES from all generated SMILES obtained from the docking task. All generative runs produced sets of SMILES with percentage validity in the high nineties.

Supplementary Table 1: Similarity task grid-search. Number of steps required to solve the similarity task by meeting one of three criteria: the generated compounds exceed an average similarity score of 0.5, the generated compounds exceed an average similarity score of 0.8 and the target compound is generated. The mean number of steps required +/- the standard deviation of three runs are reported. If a solution criterium was not met by the end of an experiment for one, two or all three of the total of three conducted experimental runs, this is denoted with a *, a ** or "Unsolved",

respectively. Run numbers in bold show the lowest mean steps required to meet each criterion for the augmentation and the learning rate searches separately, disregarding runs that was not successful in all three runs. Experiments were performed over 500 steps.

| | | Similarity >0.5 | Similarity >0.8 | Solved |
|---|---|---|---|---|
| | | Mean of Solved | Mean of Solved | Mean of Solved |
| x lr | x aug | | | |
| 1 | 1 | 398 ± 25 | unsolved | 301 ± 61 |
| 1 | 2 | 258 ± 5 | unsolved | 304 ± 73 |
| 1 | 3 | 180 ± 5 | 385 ± 60 | 187 ± 37 |
| 1 | 5 | 129 ± 7 | 288 ± 50 | 184 ± 46 |
| 1 | 10 | 97 ± 6 | 282 ± 48 | 242* |
| 1 | 20 | 64 ± 7 | 185 ± 14 | 282** |
| 2 | 1 | 259 ± 13 | unsolved | 406 ± 29 |
| 2 | 2 | 155 ± 11 | 416* | 204* |
| 2 | 3 | 140 ± 5 | 271 ± 43 | 142* |
| 2 | 5 | 93 ± 4 | 280 ± 13 | 138* |
| 2 | 10 | 63 ± 2 | 150 ± 59 | 68* |
| 2 | 20 | 48 ± 5 | 169 ± 64 | 110* |
| 3 | 1 | 210 ± 16 | 441 ± 42 | 216 ± 51 |
| 3 | 2 | 122 ± 4 | 218 ± 18 | 122 ± 10 |
| 3 | 3 | 98 ± 3 | 241 ± 39 | 138* |
| 3 | 5 | 76 ± 2 | 288 ± 86 | 178* |
| 3 | 10 | 61 ± 2 | 164 ± 41 | 82** |
| 3 | 20 | 41 ± 2 | 147 ± 50 | 48** |
| 5 | 1 | 143 ± 14 | 376* | 286* |
| 5 | 2 | 103 ± 9 | 326 ± 36 | 233** |
| 5 | 3 | 78 ± 4 | 288 ± 33 | 251** |
| 5 | 5 | 65 ± 4 | 186 ± 25 | 110** |
| 5 | 10 | 50 ± 4 | 178 ± 29 | unsolved |
| 5 | 20 | 36 ± 5 | 197 ± 30 | unsolved |
| 7 | 1 | 123 ± 4 | 401 ± 30 | 288 ± 42 |
| 7 | 2 | 86 ± 11 | 289 ± 28 | unsolved |
| 7 | 3 | 72 ± 7 | 250 ± 25 | 283** |
| 7 | 5 | 62 ± 5 | 172 ± 52 | 44** |
| 7 | 10 | 38 ± 5 | 107 ± 50 | 51* |
| 7 | 20 | 38 ± 8 | 185 ± 122 | 53** |
| 15 | 1 | 99 ± 4 | 367 ± 70 | 392** |
| 15 | 2 | 67 ± 4 | 212 ± 22 | 146** |
| 15 | 3 | 51 ± 4 | 216 ± 8 | unsolved |
| 15 | 5 | 44 ± 6 | 158* | unsolved |
| 15 | 10 | 50 ± 11 | 144 ± 46 | 56** |
| 15 | 20 | 60 ± 18 | 252 ± 59 | unsolved |

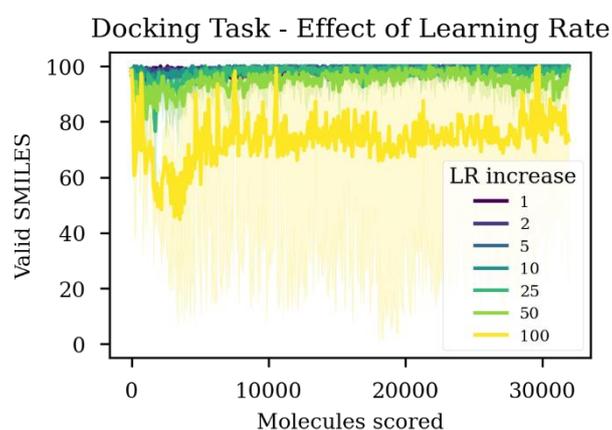

*Supplementary Figure 3.* Effect of the learning rate multiplication factor on the percentage of valid SMILES from all generated SMILES obtained from the docking task. Utilizing the highest learning rates lead to an instability in generation and lower SMILES validity.

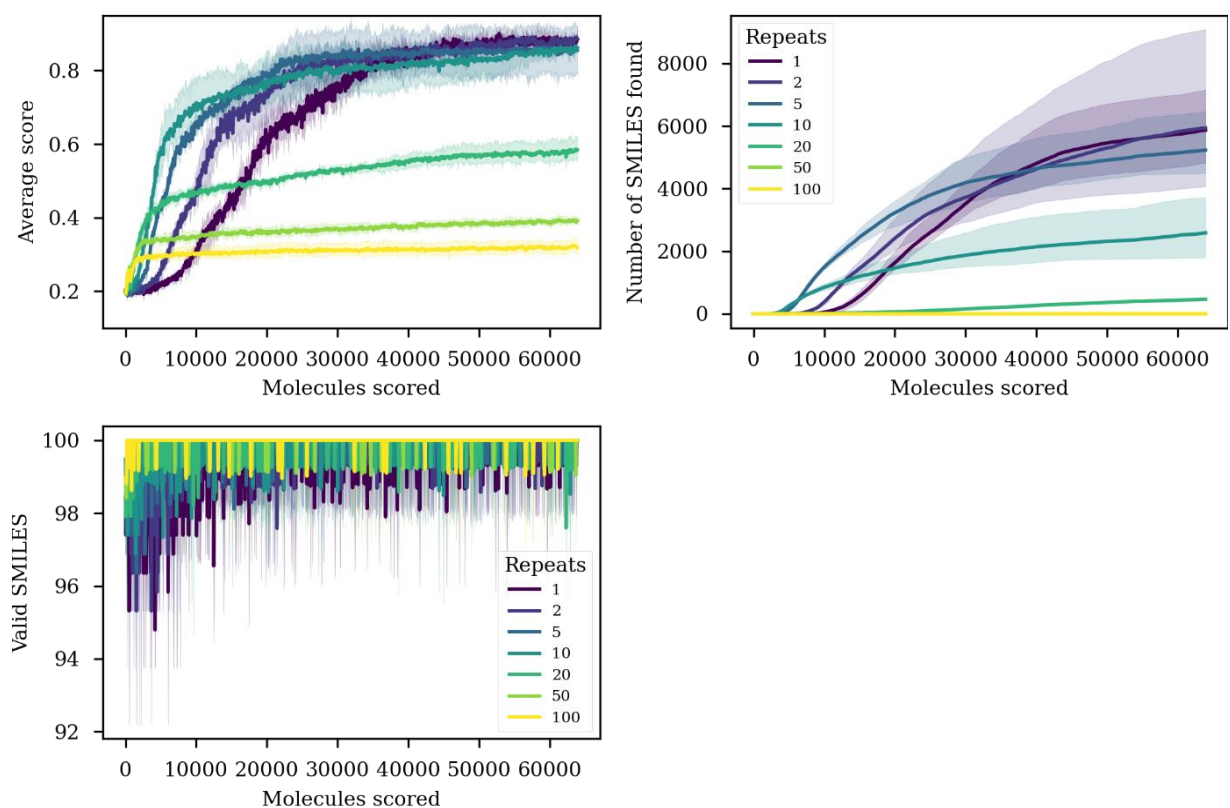

*Supplementary Figure 4. Effects of using the same SMILES sequence in the inner loop a certain number of times.*